\documentclass[12pt,english]{article}
\usepackage{times}
\usepackage[T1]{fontenc}
\usepackage[latin1]{inputenc}
\usepackage{geometry}
\usepackage{amsmath}
\usepackage{amssymb}

\usepackage{accents}
\newcommand{\f}[1]{\underaccent{\rightharpoondown}{#1}}
\newcommand{\vg}[1]{\accentset{\rightharpoonup}{#1}}
\newcommand{\ub}[1]{\underaccent{-}{#1}}

\makeatletter

\providecommand{\tabularnewline}{\\}

\usepackage{babel}
\makeatother
\begin{document}

\title{Clifford bundle formulation of BF gravity generalized to the standard
model}

\author{A. Garrett Lisi%
\thanks{alisi@hawaii.edu%
}}

\maketitle
\begin{abstract}
The structure and dynamics of the standard model and gravity are described
by a Clifford valued connection and its curvature. 
\end{abstract}
\begin{quotation}
\begin{center}{}``Everything should be made as simple as possible,
but not simpler.'' -- A.E.\end{center}
\end{quotation}

\section{Introduction}

The two most successful physical theories of the twentieth century
were quantum mechanics, culminating in Quantum Field Theory and the
standard model of particles and interactions, and General Relativity,
Einstein's geometric theory of gravitation. These two theories, refined
and verified to extraordinary accuracy, are beautiful mathematical
descriptions of the physical universe. The fact that they have been
found fundamentally incompatible stands as the greatest failure of
twentieth century science, and provides the greatest challenge at
the dawn of the twenty-first century.

Many attempts have been made to unify these two theories. The most
popular current approaches, based on string theory, extend the methods
and applications of QFT to various scenarios in a mathematically consistent
but somewhat convoluted manner only tenuously connected to the standard
model and GR. Although it is possible further development of string
theory will lead to a coherent picture -- and the development of beautiful
mathematics is certainly, itself, a noble pursuit -- several decades
of intensive research have failed to produce a single successful experimental
prediction. The one solid prediction of string theory, the existence
of super-particle partners to the existing standard model particles,
has so far failed to materialize. It is therefore reasonable to take
a step backwards, reconsider the fundamental elements of GR and the
standard model, and consider other approaches to unification following
a more conservative path.

The fundamental fields of the standard model are gauge fields, spinors,
and scalars over four dimensional spacetime. These elements have mathematical
descriptions, respectively, as fiber bundle connections, Clifford
algebra elements with anti-commuting components, and Higgs fields.
QFT calculations arising from the standard model also require the
introduction of BRST {}``ghost'' fields to properly account for
gauge degrees of freedom. General Relativity, in contrast, is about
the geometry of spacetime itself -- using a metric as well as a spin
connection. But the spinor field Lagrangian of the standard model
requires that the metric be alternatively described by a vierbein
field, also known as a frame or tetrad, as a fundamental field. Any
unified description of nature must employ all of these elements. Also,
since the naive application of QFT methods to linearized GR fail,
QFT must be generalized for a unification program to succeed.

A necessary question is how to extend QFT so it works for quantizing
GR as well as reproducing existing methods. The best current attack
on this problem is Loop Quantum Gravity \cite{rovelli book} and similar
approaches. The main modification to QFT suggested by LQG is that
quantum transition amplitudes should not be considered between field
observables at different spacetime points, but rather between boundary
surfaces, with boundary surface states described by spin networks.
This and other approaches to quantum gravity use a connection as the
fundamental field of GR. Some very interesting recent work \cite{rovelli top,smolin,freidel}
revives an idea by MacDowell and Mansouri that the $so(5)$ spin connection
may be broken into a $so(4)$ connection and vierbein, with the action
for GR given by a restricted BF action. This restricted BF action
may then be submitted to the methods of QFT, with a perturbative expansion
about the purely topological BF action.

LQG and related programs of conservatively generalizing QFT to accommodate
gravity seem, from an outsider's viewpoint, the most likely to succeed.
In effect, they constitute the first step in a program of unification
launched from the GR side rather than the particle theory side --
the final goal being to extend the geometric description of General
Relativity to encompass QFT and the standard model. The purpose of
this article is to sketch how this unification may happen at the level
of the fundamental fields, as simply as possible. In sum, the frame
and spin connection 1-forms of GR may be unified in a Clifford algebra
valued connection on a bundle, $\f{\Omega}=\f{e}+\f{\omega}$,
as in the MM method. This connection may then be incorporated in a
larger Clifford fiber with the gauge and Higgs fields of the standard
model, $\f{A}=\phi\f{e}+\f{\omega}+\f{Z}$.
In previous work, it was shown that anti-commuting Clifford fields
arise naturally from the BRST method, and the standard model fermion
multiplets may be placed in a BRST extended connection, $\tilde{\f{A}}=\phi\f{e}+\f{\omega}+\f{Z}+\psi$.
The resulting BRST extended curvature may then be used in a restricted
BF Lagrangian, giving the standard model plus gravity from a BRST
extended, Clifford algebra valued connection. Each step of this construction,
building up to the full standard model plus gravity, will be described
in detail using basic differential geometry so the reader may readily
skim, reproduce, or absorb the material.

\section{Clifford bundle connection for GR}

An {}``$n$ dimensional'' Clifford algebra fiber, $Cl_{p,q}$, is
a $2^{n}$ dimensional graded Lie algebra built from $n=p+q$ basis
vector generators, $\gamma_{\alpha}$, satisfying\begin{equation}
\gamma_{\alpha}\cdot\gamma_{\beta}=\frac{1}{2}\left(\gamma_{\alpha}\gamma_{\beta}+\gamma_{\beta}\gamma_{\alpha}\right)=\eta_{\alpha\beta}\label{eq:eta}\end{equation}
 with $\eta_{\alpha\beta}$ the generalized diagonal Minkowski metric
having $p$ positive and $q$ negative entries. Any two unequal basis
vectors anti-commute and produce a non-zero bivector, such as $\gamma_{1}\gamma_{2}=-\gamma_{2}\gamma_{1}=\gamma_{12}$.
The Clifford algebra product, equivalent to the matrix product in
a suitable representation, decomposes into symmetric and anti-symmetric
parts,\begin{align*}
A\cdot B & =\frac{1}{2}\left(AB+BA\right)\\
\left[A,B\right]=A\times B & =\frac{1}{2}\left(AB-BA\right)\end{align*}
 with this bracket operator, equivalent to the usual commutator bracket
with an added factor of $\frac{1}{2}$, extending to handle arbitrary
numbers of multivectors -- for example:\[
\left[A,B,C\right]=\frac{1}{3!}\left(ABC+BCA+CAB-ACB-CBA-BAC\right)\]
 In this way any Clifford element may be written in terms of the basis
elements as\begin{align*}
A & =A^{s}+A^{\alpha}\gamma_{\alpha}+\frac{1}{2}A^{\alpha\beta}\left[\gamma_{\alpha},\gamma_{\beta}\right]+\frac{1}{3!}A^{\alpha\beta\gamma}\left[\gamma_{\alpha},\gamma_{\beta},\gamma_{\gamma}\right]+\ldots+A^{p}\left[\gamma_{0},\ldots,\gamma_{n-1}\right]\\
 & =A^{s}+A^{\alpha}\gamma_{\alpha}+\frac{1}{2}A^{\alpha\beta}\gamma_{\alpha\beta}+\frac{1}{3!}A^{\alpha\beta\gamma}\gamma_{\alpha\beta\gamma}+\ldots+A^{p}\gamma\end{align*}
 The bivector (grade 2) part of this element is equal to $\left\langle A\right\rangle _{2}=\frac{1}{2}A^{\alpha\beta}\gamma_{\alpha\beta}=A^{2}$,
determined by the $\frac{1}{2}n(n-1)$ real coefficients, $A^{\alpha\beta}=A^{[\alpha\beta]}$,
multiplying the corresponding Lie basis elements, $T_{A}=\frac{1}{2}\gamma_{\alpha\beta}$.
The scalar (grade 0) part of a Clifford element, $\left\langle A\right\rangle =A^{s}$,
is equivalent to the trace divided by the dimension of the matrix
representation of the element; so for any Clifford elements, $\left\langle AB\right\rangle =\left\langle BA\right\rangle $.
The structure constants for the algebra may be read off identities
built from straightforward computations:\begin{align*}
\gamma_{\alpha}\times\gamma_{\beta} & =\gamma_{\alpha\beta}\\
\gamma_{\alpha}\times\gamma_{\beta\gamma} & =\eta_{\alpha\beta}\gamma_{\gamma}-\eta_{\alpha\gamma}\gamma_{\beta}\\
\gamma_{\alpha\beta}\times\gamma_{\gamma\delta} & =\eta_{\alpha\gamma}\gamma_{\beta\delta}+\eta_{\alpha\delta}\gamma_{\beta\gamma}+\eta_{\beta\gamma}\gamma_{\alpha\delta}+\eta_{\beta\delta}\gamma_{\alpha\gamma}\\
 & \vdots\end{align*}
 Equally useful identities follow from the symmetric product, such
as\[
\gamma_{\alpha\beta}\cdot\gamma_{\gamma\delta}=\left(\eta_{\alpha\delta}\eta_{\beta\gamma}-\eta_{\alpha\gamma}\eta_{\beta\delta}\right)+\gamma_{\alpha\beta\gamma\delta}\]
 giving the scalar and 4-vector parts of the symmetric product of
two bivector basis elements. Taking the scalar part of two multiplied
basis elements, both of grade $r$, gives the orthogonality relation,\[
\left\langle \gamma_{\alpha\ldots\beta}\gamma^{\gamma\ldots\delta}\right\rangle =r!\delta_{\left[\beta\right.}^{\gamma}\ldots\delta_{\left.\alpha\right]}^{\delta}\]
 with indices raised by $\eta^{\alpha\beta}$. The above identities
imply that the sub-algebra of anti-symmetric products of bivectors
of an $n$ dimensional Clifford algebra, $\textrm{spin}(n)$, is equivalent
to the algebra of anti-symmetric products of vector and bivector elements
of an $n-1$ dimensional Clifford algebra. This sub-algebra of vectors
and bivectors is nearly, but not, equivalent to the Poincare algebra
of corresponding dimension. Also, the even graded sub-algebra of an
$n$ dimensional Clifford algebra is equal to an $n-1$ dimensional
Clifford algebra.

The fundamental Clifford identity (\ref{eq:eta}), and thus the Clifford
algebra itself, is invariant under the Clifford group adjoint operation,\begin{equation}
\gamma'^{\alpha}=U\gamma U^{-1}\simeq\left(1+\frac{1}{2}C\right)\gamma^{\alpha}\left(1-\frac{1}{2}C\right)\simeq\gamma^{\alpha}+C\times\gamma^{\alpha}\label{eq:adj}\end{equation}
 in which, for infinitesimal transformations, $U\simeq1+\frac{1}{2}C$
for some {}``small'' multivector $C$. This operation gives the
form of the transition functions acting on fiber basis elements over
the base manifold.

Of special algebraic interest is the Clifford pseudo-scalar, $\gamma=\gamma_{0}\gamma_{1}\ldots\gamma_{n-1}$,
which squares to\[
\gamma\gamma=\left(-1\right)^{q}\left(-1\right)^{\frac{n\left(n+1\right)}{2}}\]
 implying $\gamma^{-1}=-\gamma$ for dimension $(1,3)$. Multiplication
by $\gamma^{-1}$ acts as the Clifford duality transformation, taking
a Clifford $r$-vector to its {}``Clifford dual'' $(n-r)$-vector,\begin{equation}
A_{r}\gamma^{-1}=\frac{1}{r!}A^{\alpha\ldots\beta}\gamma_{\alpha\ldots\beta}\gamma^{-1}=\frac{1}{r!\left(n-r\right)!}A^{\alpha\ldots\beta}\epsilon_{\alpha\ldots\beta\gamma\ldots\delta}\gamma^{\gamma\ldots\delta}\label{eq:cdual}\end{equation}
 in which $\gamma^{\gamma\ldots\delta}$ are the $(n-r)$-vector basis
elements with indices raised by $\eta$ and\[
\epsilon_{\alpha\ldots\delta}=n!\delta_{\left[\alpha\right.}^{0}\ldots\delta_{\left.\delta\right]}^{n-1}=\left\langle \gamma_{\alpha\ldots\delta}\gamma^{-1}\right\rangle \]
 is the anti-symmetric permutation symbol. The pseudo-scalar always
commutes with even graded elements, $A^{e}\gamma=\gamma A^{e}$, and
anti-commutes with odd graded elements only in even dimensions, $A^{o}\gamma=(-1)^{n+1}\gamma A^{o}$.
So for even $n$ we have $A\cdot\gamma=A^{e}\gamma$.

The covariant derivative acting on basis elements, consistent with
the transition functions, encodes how the basis vectors change as
we move around on the base manifold,\[
\f{\nabla}\gamma_{\alpha}=\f{dx^{i}}\nabla_{i}\gamma_{\alpha}=\f{\Omega}\times\gamma_{\alpha}\]
 and does not necessarily preserve the grade of the basis elements
when $\f{\Omega}$ is a Clifford valued connection 1-form
of arbitrary grade. This implies that for any section, $C$ (a Clifford
valued field over the base manifold), its covariant derivative is

\[
\f{\nabla}C=\f{\partial}C+\f{\Omega}\times C\]
 with $\f{\partial}=\f{dx^{i}}\partial_{i}$
the exterior derivative operator. The spin connection of General Relativity,
$\f{\omega}=\frac{1}{2}\f{dx^{i}}\omega_{i}{}^{\alpha\beta}\gamma_{\alpha\beta}$,
is a bivector valued 1-form, encoding how the basis vectors rotate
as we move around -- it does preserve the grade of the basis elements.
For example, an observer moving along a path $x(\tau)$ parameterized
by $\tau$ with velocity $\vg{v}=\frac{dx^{i}(\tau)}{dt}\vg{\partial}_{i}=v^{i}\vg{\partial}_{i}$
would have the basis vectors changing over them by\[
\vg{v}\f{\nabla^{\omega}}\gamma_{\alpha}=\vg{v}\f{\omega}\times\gamma_{\alpha}=v^{i}\omega_{i}{}^{\beta}{}_{\alpha}\gamma_{\mathbf{\beta}}\]
 in which the vector-form contraction rule employed above is traditionally
written less compactly as $\vg{\partial}_{i}\f{dx^{j}}=\delta_{i}^{j}={\textbf{i}}_{\vg{\partial}_{i}}\f{dx^{j}}$.
It is pedagogically useful to mark all tangent bundle vectors and
forms with accents indicating their grade so as to better identify
their nature and commutative properties. If there is a section of
the bundle, a field $C(x)$, an observer moving along a path would
see this field change over them as a function of $\tau$ to $C(x(\tau))$.
A section is said to be parallel transported by the spin connection
over their path iff\[
0=\vg{v}\f{\nabla^{\omega}}C=\vg{v}\left(\f{\partial}C+\f{\omega}\times C\right)=v^{i}\partial_{i}C+v^{i}\omega_{i}\times C=\frac{d}{d\tau}C+v^{i}\omega_{i}\times C\]
 evaluated along the path. Since the section and its derivatives are
only evaluated along the path, this extends to describe a fiber element
defined only over the path, $C(\tau)$, that is said to be parallel
transported if it satisfies the same equation.

The frame, a Clifford vector valued 1-form,\[
\f{e}=\f{dx^{i}}\left(e_{i}\right)^{\alpha}\gamma_{\alpha}=\f{dx^{i}}e_{i}=\f{e^{\alpha}}\gamma_{\alpha}\]
 encodes a metric on a base manifold of dimension $n$ through contraction
with vectors,\[
\left(\vg{v}\f{e}\right)\cdot\left(\vg{w}\f{e}\right)=v^{i}\left(\vg{\partial}_{i}\f{dx^{j}}\right)\left(e_{j}\right)^{\alpha}\gamma_{\alpha}\cdot w^{k}\left(\vg{\partial}_{k}\f{dx^{m}}\right)\left(e_{m}\right)^{\beta}\gamma_{\beta}=v^{i}w^{k}\left(\left(e_{i}\right)^{\alpha}\left(e_{k}\right)^{\beta}\eta_{\alpha\beta}\right)=v^{i}w^{k}g_{ik}\]
 and has an inverse, $\vg{e}$, such that\[
\vg{e}\f{e}=\gamma^{\beta}\left(e_{\beta}^{-1}\right)^{j}\vg{\partial}_{j}\f{dx^{i}}\left(e_{i}\right)^{\alpha}\gamma_{\alpha}=\gamma^{\beta}\left(e_{\beta}^{-1}\right)^{i}\left(e_{i}\right)^{\alpha}\gamma_{\alpha}=\gamma^{\beta}\delta_{\beta}^{\alpha}\gamma_{\alpha}=n\]

It turns out to be remarkably fruitful to consider the frame and spin
connection together in a Clifford connection,\begin{equation}
\f{\Omega}=\f{e}+\f{\omega}\label{eq:Omega}\end{equation}
 of mixed grade 1 and 2. Although this is algebraicly equivalent to
a bivector connection in one higher dimension, it seems natural to
consider this combined connection in $n=4$ over a four dimensional
base manifold.

\section{BF action for GR and equations of motion}

The geometry of a fiber bundle is described by the curvature of its
connection. Just as the connection arises as the description of how
sections change over the paths of observers traveling on the base,
the curvature continues this description to second order in path length
and determines how fiber elements change when parallel transported
around small loops.

\subsection{Parallel transport and Curvature}

Any Clifford multivector, $C$, is parallel transported along a parameterized
curve, $x(\tau)$, with velocity $\vg{v}$ iff\begin{equation}
0=v^{i}\nabla_{i}C=\vg{v}\left(\f{\partial}C+\f{\Omega}\times C\right)=\frac{d}{d\tau}C+\vg{v}\f{\Omega}\times C\label{eq:ctransp}\end{equation}
 for a specified Clifford connection. Parallel transport transforms
the multivector via a path dependent Clifford valued adjoint operator,
$U(\tau)$, as it moves along the curve,\[
C(\tau)=U(\tau)\, C(0)\, U^{-1}\]
 with initial condition $U(0)=1$. This parallel transport operator
is independent of $C$ and, from (\ref{eq:ctransp}), must satisfy
the parallel transport equation:\[
0=\frac{d}{d\tau}U+\frac{1}{2}\vg{v}\f{\Omega}U\]
 For small distances along a path, $x(\tau)=x_{0}+\varepsilon(\tau)$,
this operator may be approximated to arbitrary order. To first order\[
U(\tau)=1-\frac{1}{2}\int_{0}^{\tau}\f{dt}\,\frac{d\varepsilon^{i}}{dt}\Omega_{i}(x(t))\, U(t)\simeq1-\frac{1}{2}\varepsilon^{i}\Omega_{i}(x_{0})\]
 and to second order\begin{align}
U(\tau) & \simeq1-\frac{1}{2}\int_{0}^{\tau}\f{dt}\,\frac{d\varepsilon^{i}}{dt}\left[\Omega_{i}+\varepsilon^{j}\partial_{j}\Omega_{i}\right]\left[1-\frac{1}{2}\varepsilon^{k}\Omega_{k}\right]\nonumber \\
 & \simeq1-\frac{1}{2}\varepsilon^{i}\Omega_{i}+\frac{1}{2}\varepsilon^{ij}\left[-\partial_{j}\Omega_{i}+\frac{1}{2}\Omega_{i}\Omega_{j}\right]\label{eq:u2}\end{align}
 with the second order path dependence above defined as\[
\varepsilon^{ij}=\int_{0}^{\tau}\f{dt}\,\frac{d\varepsilon^{i}}{dt}\varepsilon^{j}\]
 Continuing the series suggests the formal expression\[
U(\tau)=\exp\left(-\frac{1}{2}\int_{0}^{\tau}\f{dt}\,\left(\vg{v}\f{\Omega}\right)\right)=\exp\left(-\frac{1}{2}\int_{0}^{\tau}\f{\Omega}\right)\]

The curvature is a geometric object determining the approximate change
in any multivector parallel transported around a small loop. A loop
may be specified by choosing two orthonormal vectors, $\vg{u}$ and
$\vg{v}$, at a point $x_{0}$ and making a square-ish path by going
$\varepsilon$ in the $\vg{u}$ direction, then $\varepsilon$ along
$\vg{v}$, $\varepsilon$ along $-\vg{u}$, then $\varepsilon$
along $-\vg{v}$ back to $x_{0}$. These four parameterized path
segments are given by\[
\varepsilon_{1}^{i}=tu^{i},\quad\varepsilon_{2}^{i}=\varepsilon u^{i}+tv^{i},\quad\varepsilon_{3}^{i}=\varepsilon u^{i}+\varepsilon v^{i}-tu^{i},\quad\varepsilon_{4}^{i}=\varepsilon v^{i}-tv^{i}\]
 and produce an anti-symmetric path dependence,\[
\varepsilon^{ij}=\int_{0}^{\varepsilon}\f{dt}\,\frac{d\varepsilon_{1}^{i}}{dt}\varepsilon_{1}^{j}+\int_{0}^{\varepsilon}\f{dt}\,\frac{d\varepsilon_{2}^{i}}{dt}\varepsilon_{2}^{j}+\int_{0}^{\varepsilon}\f{dt}\,\frac{d\varepsilon_{3}^{i}}{dt}\varepsilon_{3}^{j}+\int_{0}^{\varepsilon}\f{dt}\,\frac{d\varepsilon_{4}^{i}}{dt}\varepsilon_{4}^{j}=\varepsilon^{2}\left(v^{i}u^{j}-v^{j}u^{i}\right)\]
 implying the loop is described by a tangent 2-vector,\[
\vg{\vg{L}}=\varepsilon^{2}\vg{v}\,\vg{u}=\varepsilon^{2}v^{i}u^{j}\vg{\partial}_{i}\vg{\partial}_{j}=\frac{1}{2}\varepsilon^{ij}\vg{\partial}_{i}\vg{\partial}_{j}=\frac{1}{2}L^{ij}\vg{\partial}_{i}\vg{\partial}_{j}\]
 From (\ref{eq:u2}), the operator for parallel transport completely
around a small loop is approximately,\[
U\simeq1+\frac{1}{2}L^{ij}\left[-\partial_{j}\Omega_{i}+\frac{1}{2}\Omega_{i}\Omega_{j}\right]=1+\frac{1}{4}L^{ij}\left[\partial_{i}\Omega_{j}-\partial_{j}\Omega_{i}+\Omega_{i}\times\Omega_{j}\right]=1+\frac{1}{4}L^{ij}F_{ij}=1-\frac{1}{2}\vg{\vg{L}}\f{\f{F}}\]
 with the Clifford valued curvature 2-form coefficients here emerging
as\[
F_{ij}=\partial_{i}\Omega_{j}-\partial_{j}\Omega_{i}+\Omega_{i}\times\Omega_{j}\]
 and the index free Clifford valued curvature 2-form written as\[
\f{\f{F}}=\frac{1}{2}\f{dx^{i}}\f{dx^{j}}F_{ij}=\f{\partial}\f{\Omega}+\frac{1}{2}\f{\Omega}\times\f{\Omega}\]
 The wedge product between forms is not written since forms always
wedge, and the cross product occurs only between Clifford basis elements.
Any Clifford element, $C$, parallel transported around a small loop
is changed to\[
C\mapsto C'=UCU^{-1}\simeq C-\vg{\vg{L}}\f{\f{F}}\times C\]
 to first order in loop area.

For a bivector valued spin connection, the bivector valued Riemann
curvature 2-form is\[
\f{\f{R}}=\frac{1}{4}\f{dx^{i}}\f{dx^{j}}R_{ij}{}^{\alpha\beta}\gamma_{\alpha\beta}=\f{\partial}\f{\omega}+\frac{1}{2}\f{\omega}\times\f{\omega}\]
 for which, in components,\[
\frac{1}{2}\f{\omega}\times\f{\omega}=\frac{1}{8}\f{dx^{i}}\f{dx^{j}}\omega_{i}{}^{\alpha\beta}\omega_{j}{}^{\gamma\delta}\gamma_{\alpha\beta}\times\gamma_{\gamma\delta}=\frac{1}{2}\f{\omega}\f{\omega}=\frac{1}{2}\f{dx^{i}}\f{dx^{j}}\omega_{i}{}^{\alpha\beta}\omega_{j\alpha}{}^{\delta}\gamma_{\beta\delta}\]
 The writing of the cross between the product of any two identical
1-forms is redundant, $\f{\Omega}\times\f{\Omega}=\f{\Omega}\f{\Omega}$,
since basis 1-forms anti-commute.

The Clifford curvature of a combined vector and bivector connection
(\ref{eq:Omega}) naturally splits into Clifford vector and bivector
graded parts,\begin{align*}
\f{\f{F}} & =\f{\partial}\f{\Omega}+\frac{1}{2}\f{\Omega}\f{\Omega}\\
 & =\left(\f{\partial}\f{e}+\f{\omega}\times\f{e}\right)+\left(\f{\partial}\f{\omega}+\frac{1}{2}\f{\omega}\f{\omega}+\frac{1}{2}\f{e}\f{e}\right)\\
 & =\f{\f{T}}+\left(\f{\f{R}}+\f{\f{E}}\right)\end{align*}
 identifiable as the torsion vector valued 2-form, the Riemann curvature
bivector, and the bivector area 2-form. The curvature may also be
obtained by twice applying the covariant derivative,\[
\f{\nabla}\f{\nabla}C=\f{\f{F}}\times C\]

\subsection{Action}

Over a four dimensional base manifold a restricted BF action equivalent
to that of General Relativity may be written with Clifford $n=4$
as\begin{equation}
S=\int\ub{L}=\int\left\langle \f{\f{B}}\f{\f{F}}-\frac{1}{2}\f{\f{B}}\f{\f{B}}\gamma\right\rangle \label{eq:Oaction}\end{equation}
 employing the Clifford pseudo-scalar, $\gamma=(-1)^{q}\gamma^{-1}$,
a new vector and bivector valued 2-form variable, $\f{\f{B}}$,
and using an under-bar to designate forms of large or arbitrary grade.
Since Clifford basis orthogonality implies the scalar part of two
multiplied Clifford elements splits into terms of multiplied equal
graded parts, this Lagrangian splits into\[
\ub{L}=\left\langle \f{\f{B^{1}}}\f{\f{T}}+\f{\f{B^{2}}}\left(\f{\f{R}}+\f{\f{E}}-\frac{1}{2}\f{\f{B^{2}}}\gamma\right)\right\rangle \]
 Varying the action with respect to $\f{\f{B}}$,
the vector part, $\f{\f{B^{1}}}$, is
a Lagrange multiplier enforcing zero torsion, $\f{\f{T}}=0$,
and the bivector part gives the equation\begin{equation}
\f{\f{B^{2}}}=\left(\f{\f{R}}+\f{\f{E}}\right)\gamma^{-1}=\f{\f{F^{2}}}\gamma^{-1}\label{eq:Beom}\end{equation}
 in which $\f{\f{F^{2}}}$ is the bivector
part of the curvature. Plugging this back in gives the Lagrangian
purely in terms of the curvature,\begin{equation}
\ub{L}=\frac{1}{2}\left\langle \left(\f{\f{R}}+\f{\f{E}}\right)\left(\f{\f{R}}+\f{\f{E}}\right)\gamma^{-1}\right\rangle =\frac{1}{2}\left\langle \f{\f{F}}\f{\f{F}}\gamma^{-1}\right\rangle \label{eq:ffg}\end{equation}
 Since the Clifford pseudo-scalar commutes with even graded Clifford
elements, basis 1-forms anti-commute, and the exterior derivative
is nilpotent, the Riemann squared term in (\ref{eq:ffg}) is a Chern-Simons
boundary term,\[
\left\langle \f{\f{R}}\f{\f{R}}\gamma^{-1}\right\rangle =\left\langle \left(\f{\partial}\f{\omega}+\frac{1}{2}\f{\omega}\f{\omega}\right)\left(\f{\partial}\f{\omega}+\frac{1}{2}\f{\omega}\f{\omega}\right)\gamma^{-1}\right\rangle =\f{\partial}\left\langle \left(\f{\omega}\left(\f{\partial}\f{\omega}\right)+\frac{1}{3}\f{\omega}\f{\omega}\f{\omega}\right)\gamma^{-1}\right\rangle \]
 So the Lagrangian is equivalent, up to this boundary term, to the
Lagrangian for Einstein-Cartan gravity including a cosmological constant
term,\[
\ub{L^{\textrm{eff}}}=\frac{1}{2}\left\langle \left(\f{\f{E}}\f{\f{R}}+\f{\f{R}}\f{\f{E}}+\f{\f{E}}\f{\f{E}}\right)\gamma^{-1}\right\rangle =\frac{1}{2}\left\langle \left(\f{e}\f{e}\f{\f{R}}+\frac{1}{4}\f{e}\f{e}\f{e}\f{e}\right)\gamma^{-1}\right\rangle =\ub{e}\frac{1}{2}\left\langle R+6\right\rangle \]
 in which $\ub{e}=\frac{1}{4!}\left\langle \f{e}\f{e}\f{e}\f{e}\gamma^{-1}\right\rangle =\ub{d^{4}x}\left|e\right|$
is the volume 4-form and $R=\left\langle \vg{e}\vg{e}\f{\f{R}}\right\rangle $
is the curvature scalar. The appearance of the cosmological constant
is made explicit by the appropriate vierbein scaling -- rescaling
$\f{e}$ to $\sqrt{\Lambda}\f{e}$. The
equivalence of terms written in Clifford valued form notation to those
written in components is seen by expanding in basis elements and employing
the orthogonality rules. Although the methods and action described
so far are equivalent to those of MacDowell-Mansouri and others, the
formulation here allows the action to be written naturally in terms
of the Clifford dual and does not require symmetry breaking to step
down from $so(5)$. Also, the Clifford algebra approach easily generalizes
to additional interesting systems, such as Clifford algebras and sub-algebras
of different signatures and dimensions. As in the MM approach, the
main observation is that the dynamics of GR may be described purely
in terms of a connection, without needing a metric on the base manifold.
Viewed in this light, the scale of the vierbein should properly be
interpreted as a Higgs field -- an idea that will pop up again later.

\subsection{Equations of motion}

The first equation of motion is the vanishing of the torsion,\begin{equation}
0=\f{\f{T}}=\f{\partial}\f{e}+\f{\omega}\times\f{e}\label{eq:eom1}\end{equation}
 which, for $n=4$, may be solved explicitly for the spin connection
in terms of the exterior derivative and inverse of the vierbein,\[
\f{\omega}=-\vg{e}\times\left(\f{\partial}\f{e}\right)+\frac{1}{4}\left(\vg{e}\times\vg{e}\right)\left(\f{e}\cdot\left(\f{\partial}\f{e}\right)\right)\]

Varying the action (\ref{eq:Oaction}) with respect to $\f{\Omega}$
gives\[
\delta S=\int\left\langle \f{\f{B}}\left(\f{\partial}\f{\delta\Omega}+\frac{1}{2}\f{\delta\Omega}\f{\Omega}+\frac{1}{2}\f{\Omega}\f{\delta\Omega}\right)\right\rangle =\int\left\langle \f{\delta\Omega}\left(\f{\partial}\f{\f{B}}+\frac{1}{2}\f{\Omega}\f{\f{B}}-\frac{1}{2}\f{\f{B}}\f{\Omega}\right)+\f{\partial}\left(\f{\f{B}}\f{\delta\Omega}\right)\right\rangle \]
 and hence the second equation of motion,\[
0=\f{\partial}\f{\f{B}}+\f{\Omega}\times\f{\f{B}}=\f{\nabla}\f{\f{B}}\]
 which includes the odd and even graded parts,\begin{align*}
0 & =\f{\partial}\f{\f{B^{1}}}+\f{\omega}\times\f{\f{B^{1}}}+\f{e}\times\f{\f{B^{2}}}\\
0 & =\f{\partial}\f{\f{B^{2}}}+\f{\omega}\times\f{\f{B^{2}}}+\f{e}\times\f{\f{B^{1}}}\end{align*}
 Incorporating the first equation of motion (\ref{eq:Beom}) and the
{}``second'' Bianchi identity,\[
\f{\nabla^{\omega}}\f{\f{R}}=\f{\partial}\f{\f{R}}+\f{\omega}\times\f{\f{R}}=\f{\partial}\left(\f{\partial}\f{\omega}+\frac{1}{2}\f{\omega}\f{\omega}\right)+\f{\omega}\times\left(\f{\partial}\f{\omega}+\frac{1}{2}\f{\omega}\f{\omega}\right)=0\]
 these become\begin{align*}
0 & =\f{\partial}\f{\f{B^{1}}}+\f{\omega}\times\f{\f{B^{1}}}+\f{e}\times\left(\left(\f{\f{R}}+\f{\f{E}}\right)\gamma^{-1}\right)\\
0 & =\f{\partial}\f{\f{E}}\gamma^{-1}+\f{\omega}\times\f{\f{E}}\gamma^{-1}+\f{e}\times\f{\f{B^{1}}}\end{align*}
 Using the vanishing torsion, the last equation becomes $\f{\f{B^{1}}}=0$
and the only remaining equation is\begin{equation}
0=\f{e}\cdot\left(\f{\f{R}}+\f{\f{E}}\right)\label{eq:eom2}\end{equation}
 Einstein's equation, where $\f{e}\gamma=-\gamma\f{e}$
has been used, presuming even $n$.

The equations of motion may alternatively be obtained by varying $\f{\Omega}$
in (\ref{eq:ffg}) to get\begin{equation}
0=\f{\nabla}\left(\f{\f{F}}\cdot\gamma^{-1}\right)\label{eq:eom}\end{equation}
 Combining with the Clifford Bianchi identity, $\f{\nabla}\f{\f{F}}=0$,
and breaking into graded parts gives vanishing torsion, Einstein's
equation, and the {}``first'' Bianchi identity, $\f{e}\times\f{\f{R}}=0$.

\subsection{Gauge symmetry}

Under a gauge transformation (\ref{eq:adj}), parameterized by Clifford
element $U(x)\simeq1+\frac{1}{2}C(x)$, the connection transforms
to\[
\f{\Omega'}=U\f{\Omega}U^{-1}+2U\left(\f{\partial}U^{-1}\right)\simeq\f{\Omega}+\frac{1}{2}C\f{\Omega}-\frac{1}{2}\f{\Omega}C-\f{\partial}C=\f{\Omega}-\f{\nabla}C\]
 which may be written as\begin{equation}
\delta_{C}\f{\Omega}=-\f{\nabla}C\label{eq:gtransf}\end{equation}
 and the curvature transforms to\[
\f{\f{F'}}=\f{\partial}\f{\Omega'}+\frac{1}{2}\f{\Omega'}\f{\Omega'}=U\f{\f{F}}U^{-1}\simeq\f{\f{F}}+C\times\f{\f{F}}\]
 giving $\delta_{C}\f{\f{F}}=C\times\f{\f{F}}$.
This produces a transformation of the Lagrangian (\ref{eq:ffg}) to\[
\ub{L'}=\frac{1}{2}\left\langle \f{\f{F'}}\f{\f{F'}}\gamma^{-1}\right\rangle =\frac{1}{2}\left\langle \f{\f{F}}\f{\f{F}}U^{-1}\gamma^{-1}U\right\rangle \simeq\ub{L}-\frac{1}{2}\left\langle \f{\f{F}}\f{\f{F}}\left(C\times\gamma^{-1}\right)\right\rangle =\ub{L}-\left\langle \gamma^{-1}\left(\f{\f{T}}\cdot\f{\f{F^{2}}}\right)C^{1}\right\rangle \]
 in which $C^{1}$ is the Clifford vector part of $C=C^{1}+C^{2}$.
The gauge transformation decomposes into vector and bivector parts,\begin{align*}
\delta_{C}\f{e} & =-\f{\partial}C^{1}-\f{\omega}\times C^{1}-\f{e}\times C^{2}\\
\delta_{C}\f{\omega} & =-\f{\partial}C^{2}-\f{\omega}\times C^{2}-\f{e}\times C^{1}\end{align*}
 $C^{2}$ parameterizing Lorentz transformations and $C^{1}$ related
to diffeomorphisms. For a transformation to be a symmetry, the action
must be invariant up to a boundary term under the transformation.
In the space of all possible connections, $\f{\Omega}$,
the space of solutions to the equations of motion, (\ref{eq:eom}),
form a subspace referred to as the {}``shell.'' An equation that
holds only when the equations of motion are enforced is said to be
true {}``on shell,'' and one that holds even when they are not enforced
is true {}``off shell.'' Since $\delta_{C^{2}}\ub{L}=0$,
the gauge transformation parameterized by an arbitrary bivector, $C=C^{2}=\Sigma$,
is an {}``off shell'' symmetry -- giving zero variation to the Lagrangian
even if the equations of motion are not enforced. The gauge transformation
generated by arbitrary $C=C^{1}$ is an {}``on shell'' symmetry
-- giving $\delta_{C^{1}}\ub{L}=0$ only when $\f{\f{T}}=0$.
However, the gauge transformation parameterized by $C=C^{1}$ is an
{}``off shell'' symmetry if it is constrained such that the change
in the Lagrangian is exact,\begin{equation}
\delta_{C^{1}}\ub{L}=-\left\langle \gamma^{-1}\left(\f{\f{T}}\cdot\f{\f{F^{2}}}\right)C^{1}\right\rangle =\f{\partial}\ub{b}\label{eq:gcon}\end{equation}
 with $\ub{b}$ some scalar valued 3-form. This space of constrained
gauge transformations is equivalent to the space of diffeomorphisms.
If this constraint is not imposed, and arbitrary {}``on shell''
gauge transformations are allowed, it would be possible to make the
frame vanish, $\f{e'}=\f{e}+\delta_{C}\f{e}=0,$
by making a gauge transformation via $C=x^{i}e_{i}$.

A diffeomorphism consists of moving the fields over the base manifold
along an arbitrary flow field, $\vg{\xi}(x)$. The transformation
is given by the Lie derivative, applying to any geometrically defined
object:\begin{align*}
\delta_{\xi}\f{\Omega} & =\pounds_{\vg{\xi}}\,\f{\Omega}=\vg{\xi}\left(\f{\partial}\f{\Omega}\right)+\f{\partial}\left(\vg{\xi}\f{\Omega}\right)\\
\delta_{\xi}\f{\f{F}} & =\pounds_{\vg{\xi}}\,\f{\f{F}}=\vg{\xi}\left(\f{\partial}\f{\f{F}}\right)+\f{\partial}\left(\vg{\xi}\f{\f{F}}\right)\end{align*}
 The Lagrangian changes under a diffeomorphism by an exact term since
5-forms are zero over a four dimensional base manifold,\begin{equation}
\delta_{\xi}\ub{L}=\pounds_{\vg{\xi}}\,\ub{L}=\f{\partial}\left(\vg{\xi}\ub{L}\right)=\f{\partial}\ub{b}^{\xi}\label{eq:dxl}\end{equation}
 and diffeomorphisms are thus an {}``off shell'' symmetry. The gauge
transformation corresponding to a diffeomorphism may be found by solving
$\delta_{C}\f{\Omega}=\delta_{\xi}\f{\Omega}$
for $C=C^{\xi}$ given any $\vg{\xi}$ -- i.e. by finding the solution
to\begin{equation}
-\f{\nabla}C=\pounds_{\vg{\xi}}\,\f{\Omega}=\vg{\xi}\f{\f{F}}+\f{\nabla}\left(\vg{\xi}\f{\Omega}\right)\label{eq:diffcon}\end{equation}
 If the solution is split into $C^{\xi}=-\vg{\xi}\f{\Omega}+C'$,
this equation simplifies further to solving\[
\f{\nabla}C'=-\vg{\xi}\f{\f{F}}\]
 for $C'$ to get the correct gauge transformation corresponding to
any diffeomorphism.

\subsection{Hamiltonian formulation (an optional interlude)}

The variational formulation and derivation of equations of motion
may be recast in a canonical Hamiltonian framework. A functional derivative
with respect to an arbitrary p-form, $\frac{\partial}{\partial\ub{A}}$,
may be defined so the chain rule for the exterior derivative works
as, for example,\[
\f{\partial}\ub{G}(A,B)=\left(\f{\partial}\ub{A}\right)\left(\frac{\partial}{\partial\ub{A}}\ub{G}\right)+\left(\f{\partial}\ub{B}\right)\left(\frac{\partial}{\partial\ub{B}}\ub{G}\right)\]
 for an arbitrary Clifford valued form functional, $\ub{G}$,
of arbitrary Clifford valued forms, $\ub{A}$ and $\ub{B}$.
Although the above formula provides the most practical working definition
for extracting an arbitrary functional derivative, we may also define
the derivative with respect to a Clifford r-vector valued p-form in
terms of coordinate and Clifford basis elements as\[
\frac{\partial}{\partial\ub{A}}\ub{G}=p!r!\vg{\partial}_{j}\ldots\vg{\partial}_{i}\gamma^{\beta\ldots\alpha}\frac{\partial}{\partial A_{i\ldots j}{}^{\alpha\ldots\beta}}\ub{G}\]
 For an action and scalar valued Lagrangian 4-form functional of a
connection 1-form, $\ub{L}(A,\partial A)$, the system may be
cast in {}``first order'' form by defining the Clifford valued momentum
2-form,\[
\f{\f{B}}=\frac{\partial}{\partial\left(\f{\partial}\f{A}\right)}\ub{L}\]
 and scalar valued Hamiltonian 4-form,\[
\ub{H}(A,B)=\left\langle \f{\f{B}}\f{\partial}\f{A}\right\rangle -\ub{L}\]
 with $\f{\partial}\f{A}$ written in
terms of $\f{\f{B}}$. The variation of
the action in terms of these variables is\[
\delta S=\delta\int\ub{L}=\delta\int\left\langle \f{\f{B}}\f{\partial}\f{A}-\ub{H}\right\rangle =\int\left\langle \left(\delta\f{\f{B}}\right)\left(\f{\partial}\f{A}-\frac{\partial}{\partial\f{\f{B}}}\ub{H}\right)+\left(\delta\f{A}\right)\left(\f{\partial}\f{\f{B}}-\frac{\partial}{\partial\f{A}}\ub{H}\right)+\f{\partial}\left(\f{\f{B}}\left(\delta\f{A}\right)\right)\right\rangle \]

The restricted BF action for gravity (\ref{eq:Oaction}) is already
in Hamiltonian form, with connection $\f{\Omega}$ and
momentum $\f{\f{B}}$ the canonical variables,
and the Hamiltonian \begin{equation}
\ub{H}=\left\langle -\frac{1}{2}\f{\f{B}}\f{\Omega}\f{\Omega}+\frac{1}{2}\f{\f{B}}\f{\f{B}}\gamma\right\rangle \label{eq:Hgr}\end{equation}
 A restricted BF action may be used in a perturbative expansion carried
out around the solutions of pure BF theory, $\f{\f{F}}=0$,
corresponding to Hamiltonian perturbation around $\ub{H}_{0}=\left\langle -\frac{1}{2}\f{\f{B}}\f{\Omega}\f{\Omega}\right\rangle $.
The canonical equations of motion for BF restricted gravity are a
re-assemblage of the equations of motion into\begin{align}
\f{\partial}\f{\Omega} & =\frac{\partial}{\partial\f{\f{B}}}\ub{H}=-\frac{1}{2}\f{\Omega}\f{\Omega}+\f{\f{B}}\cdot\gamma\nonumber \\
\f{\partial}\f{\f{B}} & =\frac{\partial}{\partial\f{\Omega}}\ub{H}=-\f{\Omega}\times\f{\f{B}}\label{eq:Heom}\end{align}
 This description of motion in terms of exterior derivatives is particularly
well adapted to describe flows through boundary surfaces. To recover
the traditional coordinate based description, we need only write out
the forms in terms of coordinates including time, $x^{0}$, such as\[
\f{\partial}\f{e}=\left(\f{dx^{0}}\partial_{0}+\f{dx^{a}}\partial_{a}\right)\left(\f{dx^{0}}e_{0}+\f{dx^{b}}e_{b}\right)=\f{dx^{0}}\f{dx^{a}}\left(\partial_{0}e_{a}-\partial_{a}e_{0}\right)+\f{dx^{a}}\f{dx^{b}}\partial_{a}e_{b}\]
 We may alternatively decompose the forms with respect to a level
surface. For some scalar valued function over the base manifold, $t(x)$,
any p-form may be split into parts parallel and perpendicular to a
surface of constant $t$ as\[
\ub{A}=\ub{A}^{\Vert}+\ub{A}^{\bot}=\vg{t}\left(\f{dt}\ub{A}\right)+\f{dt}\left(\vg{t}\ub{A}\right)\]
 with $\f{dt}=\f{dx^{i}}\partial_{i}t$
and its dual vector, $\vg{t}$, satisfying $\vg{t}\f{dt}=1$.
Or, as a third alternative, we could decompose the equations of motion
via the Lie derivative,\[
\pounds_{\vg{t}}\,\ub{A}=\vg{t}\left(\f{\partial}\ub{A}\right)+\f{\partial}\left(\vg{t}\ub{A}\right)\]

The canonical equations of motion may also be obtained by defining
a Poisson-like bracket operator,\[
\left\{ \ub{F},\ub{G}\right\} =\left(\frac{\partial}{\partial\f{A}}\ub{F}\right)\left(\frac{\partial}{\partial\f{\f{B}}}\ub{G}\right)+\left(-1\right)^{o(\ub{F})}\left(\frac{\partial}{\partial\f{\f{B}}}\ub{F}\right)\left(\frac{\partial}{\partial\f{A}}\ub{G}\right)\]
 in which $o(\ub{F})$ is the form order. We may also write
$\delta_{\ub{F}}\ub{G}=\{\ub{F},\ub{G}\}$
for {}``the canonical transformation of $\ub{G}$ generated
by $\ub{F}$.'' Since $\f{A}$ is a 1-form, this
Poisson-like bracket is not necessarily anti-symmetric and should
only be considered a calculational convenience, though it does satisfy
$\{\f{A},\f{\f{B}}\}=-\{\f{\f{B}},\f{A}\}=1$.
The Hamiltonian is the generator of the part of the exterior derivative
dependent on the canonical variables. For some functional, $\ub{G}(A,B,C)$,
of the canonical variables and a parameterizing field, $C(x)$, the
exterior derivative is \begin{equation}
\f{\partial}\ub{G}=\left(\f{\partial}C\right)\left(\frac{\partial}{\partial C}\ub{G}\right)+\left(\f{\partial}\f{\f{B}}\right)\left(\frac{\partial}{\partial\f{\f{B}}}\ub{G}\right)+\left(\f{\partial}\f{A}\right)\left(\frac{\partial}{\partial\f{A}}\ub{G}\right)=\left(\f{\partial}C\right)\left(\frac{\partial}{\partial C}\ub{G}\right)+\left\{ \ub{H},\ub{G}\right\} \label{eq:extdiff}\end{equation}
 when evaluated {}``on shell'' -- when the equations of motion (\ref{eq:Heom})
are satisfied. Canonical transformations are made by choosing a generating
functional. A particularly useful class of these are scalar valued
3-form functionals, $\ub{G}(A,B,C)$, parameterized by $C(x)$.
Such functionals produce a canonical transformation,\begin{align*}
\delta_{C}\f{A}=\left\{ \ub{G},\,\f{A}\right\}  & =-\frac{\partial}{\partial\f{\f{B}}}\ub{G}\\
\delta_{C}\f{\f{B}}=\left\{ \ub{G},\,\f{\f{B}}\right\}  & =\frac{\partial}{\partial\f{A}}\ub{G}\end{align*}
 and a variation in the action,\begin{align*}
\delta_{C}S & =\int\left\langle \left(\delta_{C}\f{\f{B}}\right)\left(\f{\partial}\f{A}-\frac{\partial}{\partial\f{\f{B}}}\ub{H}\right)+\left(\delta_{C}\f{A}\right)\left(\f{\partial}\f{\f{B}}-\frac{\partial}{\partial\f{A}}\ub{H}\right)+\f{\partial}\left(\f{\f{B}}\delta_{C}\f{A}\right)\right\rangle \\
 & =\int\left\langle \left(\frac{\partial}{\partial\f{A}}\ub{G}\right)\left(\f{\partial}\f{A}-\frac{\partial}{\partial\f{\f{B}}}\ub{H}\right)-\left(\frac{\partial}{\partial\f{\f{B}}}\ub{G}\right)\left(\f{\partial}\f{\f{B}}-\frac{\partial}{\partial\f{A}}\ub{H}\right)-\f{\partial}\left(\f{\f{B}}\frac{\partial}{\partial\f{\f{B}}}\ub{G}\right)\right\rangle \\
 & =\int\left\langle \f{\partial}\ub{G}-\left(\f{\partial}C\right)\left(\frac{\partial}{\partial C}\ub{G}\right)-\left\{ \ub{H},\,\ub{G}\right\} -\f{\partial}\left(\f{\f{B}}\frac{\partial}{\partial\f{\f{B}}}\ub{G}\right)\right\rangle \end{align*}
 The transformation is a symmetry of the action iff the variation
of the action results in a boundary term,\[
\delta_{C}\ub{L}=\f{\partial}\ub{b}\]
 which happens iff $\ub{G}$ satisfies\begin{equation}
\left\langle \left(\f{\partial}C\right)\left(\frac{\partial}{\partial C}\ub{G}\right)+\left\{ \ub{H},\,\ub{G}\right\} \right\rangle =\f{\partial}\ub{g}\label{eq:symreq}\end{equation}
 for some $\ub{g}$, in which case\begin{equation}
\ub{b}=\ub{G}-\ub{g}-\left\langle \f{\f{B}}\frac{\partial}{\partial\f{\f{B}}}\ub{G}\right\rangle \label{eq:bGg}\end{equation}
 Furthermore, when the equations of motion are satisfied (on shell),
there is a $C$ dependent conserved current related to the symmetry,
$\ub{J}=\ub{G}-\ub{g}$, which by (\ref{eq:extdiff})
and (\ref{eq:symreq}) satisfies\[
\f{\partial}\ub{J}=\f{\partial}\left(\ub{G}-\ub{g}\right)=\left\langle \left(\f{\partial}C\right)\left(\frac{\partial}{\partial C}\ub{G}\right)+\left\{ \ub{H},\,\ub{G}\right\} \right\rangle -\f{\partial}\ub{g}=0\]
 for all choices of $C$.

The generator corresponding to the gauge transformation for gravity
(\ref{eq:gtransf}) is\[
\ub{G}=\left\langle \f{\f{B}}\f{\nabla}C\right\rangle =\left\langle \f{\f{B}}\left(\f{\partial}C+\frac{1}{2}\f{\Omega}C-\frac{1}{2}C\f{\Omega}\right)\right\rangle \]
 with Clifford vector and bivector valued gauge parameter field, $C$.
This generator produces transformations by\begin{align*}
\delta_{C}\f{\Omega} & =-\frac{\partial}{\partial\f{\f{B}}}\ub{G}=-\f{\nabla}C\\
\delta_{C}\f{\f{B}} & =\frac{\partial}{\partial\f{\Omega}}\ub{G}=\frac{1}{2}C\f{\f{B}}-\frac{1}{2}\f{\f{B}}C=C\times\f{\f{B}}\end{align*}
 familiar as the Clifford adjoint gauge transformation. It also gives,
through (\ref{eq:bGg}), $\ub{g}=-\ub{b}$ and satisfies
(\ref{eq:symreq}) on shell. The generator corresponding to Lorentz
transformations, parameterized by a bivector, $C=C^{2}=\Sigma$, is\[
\ub{G}^{\Sigma}=\left\langle \f{\f{B}}\f{\nabla}\Sigma\right\rangle \]
 which generates a symmetry transformation satisfying (\ref{eq:symreq})
off shell and giving $\ub{g}^{\Sigma}=-\ub{b}^{\Sigma}=0$.
The related conserved current is\[
\ub{J}^{\Sigma}=\ub{G}^{\Sigma}=\left\langle \f{\f{B}}\f{\nabla}\Sigma\right\rangle =\left\langle \f{\f{B}}\left(\f{\partial}\Sigma+\f{\Omega}\times\Sigma\right)\right\rangle =\left\langle \Sigma\left(\f{\partial}\f{\f{B}}+\f{\Omega}\times\f{\f{B}}\right)\right\rangle +\f{\partial}\left\langle \f{\f{B}}\Sigma\right\rangle \]
 for any $\Sigma$. When considered off shell, $\f{\partial}\ub{J}^{\Sigma}=0$
implies the constraint equation associated to the Lorentz symmetry,\[
0=\left\langle \f{\nabla}\f{\f{B}}\right\rangle _{2}=\left(\f{\partial}\f{\f{B^{2}}}+\f{\omega}\times\f{\f{B^{2}}}+\f{e}\times\f{\f{B^{1}}}\right)\]
 equivalent to part of one of the equations of motion (\ref{eq:Heom}).

For a diffeomorphism, the gauge parameter is constrained to solve
(\ref{eq:diffcon}), so the generator, parameterized by $\vg{\xi}$,
is\[
\ub{G}^{\xi}=\left\langle \f{\f{B}}\f{\nabla}C^{\xi}\right\rangle =-\left\langle \f{\f{B}}\pounds_{\vg{\xi}}\,\f{\Omega}\right\rangle \]
 giving the canonical transformations\begin{align*}
\delta_{\xi}\f{\Omega} & =-\frac{\partial}{\partial\f{\f{B}}}\ub{G}^{\xi}=-\f{\nabla}C^{\xi}=\pounds_{\vg{\xi}}\,\f{\Omega}\\
\delta_{\xi}\f{\f{B}} & =\frac{\partial}{\partial\f{\Omega}}\ub{G}^{\xi}=C^{\xi}\times\f{\f{B}}=\pounds_{\vg{\xi}}\,\f{\f{B}}\end{align*}
 and, from (\ref{eq:dxl}),\[
\ub{g}^{\xi}=-\ub{b}^{\xi}=-\vg{\xi}\ub{L}=-\vg{\xi}\left\langle \f{\f{B}}\f{\partial}\f{\Omega}-\ub{H}\right\rangle \]
 and the conserved current associated to diffeomorphisms,\[
\ub{J}^{\xi}=\ub{G}^{\xi}-\ub{g}^{\xi}=-\left\langle \f{\f{B}}\pounds_{\vg{\xi}}\,\f{\Omega}\right\rangle +\vg{\xi}\left\langle \f{\f{B}}\f{\partial}\f{\Omega}-\ub{H}\right\rangle =-\vg{\xi}\ub{H}+\left\langle \left(\f{\partial}\f{\f{B}}\right)\left(\vg{\xi}\f{\Omega}\right)+\left(\vg{\xi}\f{\f{B}}\right)\left(\f{\partial}\f{\Omega}\right)\right\rangle -\f{\partial}\left\langle \f{\f{B}}\left(\vg{\xi}\f{\Omega}\right)\right\rangle \]

\section{$Cl$ representations}

A Clifford algebra of dimension $n$ has a faithful representation
in the complex matrices, $GL(2^{[n/2]},\mathbb{C})$, with the Clifford
product isomorphic to matrix multiplication. This corresponds to the
traditional use of Pauli and Dirac matrices to represent the basis
vectors, $\gamma_{\mu}$. Clifford algebra elements may also be represented
as matrices of reals, complex numbers, or quaternions -- depending
on signature. The Clifford algebra $Cl_{0,2}=\mathbb{H}$, equivalent
to the algebra of quaternions, is generated by the $2\times2$ complex,
anti-Hermitian, Clifford grade 1 basis vectors with off-diagonal elements,\[
K=q_{1}=i\sigma_{1}=\left[\begin{array}{cc}
0 & i\\
i & 0\end{array}\right]\qquad J=q_{2}=i\sigma_{2}=\left[\begin{array}{cc}
0 & 1\\
-1 & 0\end{array}\right]\]
 Their products generate the grade 0 Hermitian scalar and the grade
2 anti-Hermitian pseudo-scalar,\[
1=q_{0}=i\sigma_{0}=\left[\begin{array}{cc}
1 & 0\\
0 & 1\end{array}\right]\qquad I=q_{3}=i\sigma_{3}=\left[\begin{array}{cc}
i & 0\\
0 & -i\end{array}\right]\]
 completing the description of the four elements of $Cl_{0,2}$ and
$\mathbb{H}$ in terms of matrices. The Clifford basis elements, represented
by these Pauli matrices, satisfy the commutation relations,\[
q_{\pi}\times q_{\rho}=i\sigma_{\pi}\times i\sigma_{\rho}=-\epsilon_{\pi\rho\sigma}i\sigma_{\sigma}=\epsilon_{\pi\rho}{}^{\sigma}q_{\sigma}\]
 with these Greek indices running from $1$ to $3$. An arbitrary
quaternion is encoded by four real or two complex numbers,\begin{equation}
h=h^{\mu}q_{\mu}=\left[\begin{array}{cc}
h^{0}+ih^{3} & h^{2}+ih^{1}\\
-h^{2}+ih^{1} & h^{0}-ih^{3}\end{array}\right]=\left[\begin{array}{cc}
h_{\uparrow} & -h_{\downarrow}^{*}\\
h_{\downarrow} & h_{\uparrow}^{*}\end{array}\right]\label{eq:quat}\end{equation}
 with a star denoting complex conjugation.

A Clifford algebra relevant to spacetime, $Cl_{1,3}$, may be generated
by the four $4\times4$ complex Hermitian or anti-Hermitian, Clifford
grade 1 basis vectors with off-diagonal blocks,\begin{equation}
\gamma_{0}=\left[\begin{array}{cc}
0 & i\sigma_{0}\\
i\sigma_{0} & 0\end{array}\right]=\left[\begin{array}{cc}
0 & q_{0}\\
q_{0} & 0\end{array}\right]=\left[\begin{array}{cc}
0 & 1\\
1 & 0\end{array}\right]\qquad\gamma_{\pi}=\left[\begin{array}{cc}
0 & \sigma_{\pi}\\
-\sigma_{\pi} & 0\end{array}\right]=\left[\begin{array}{cc}
0 & -iq_{\pi}\\
iq_{\pi} & 0\end{array}\right]\label{eq:weylrep}\end{equation}
 the {}``chiral'' Weyl representation. These give the six bivector
basis elements,\[
\gamma_{0\pi}=\left[\begin{array}{cc}
-\sigma_{\pi} & 0\\
0 & \sigma_{\pi}\end{array}\right]=\left[\begin{array}{cc}
iq_{\pi} & 0\\
0 & -iq_{\pi}\end{array}\right]\qquad\gamma_{\pi\rho}=\left[\begin{array}{cc}
-q_{\pi}q_{\rho} & 0\\
0 & -q_{\pi}q_{\rho}\end{array}\right]=-\epsilon_{\pi\rho}{}^{\sigma}\left[\begin{array}{cc}
q_{\sigma} & 0\\
0 & q_{\sigma}\end{array}\right]\]
 and pseudo-scalar,\begin{equation}
\gamma=\gamma_{0}\gamma_{1}\gamma_{2}\gamma_{3}=\left[\begin{array}{cc}
-iq_{0} & 0\\
0 & iq_{0}\end{array}\right]=\left[\begin{array}{cc}
-i & 0\\
0 & i\end{array}\right]=-\gamma^{-1}\label{eq:pseudo}\end{equation}
 The catalog of sixteen basis elements for $Cl_{1,3}$ is completed
by ${\textbf{1}}$ and the basis trivectors,\[
-\gamma_{1}\gamma_{2}\gamma_{3}=\gamma_{0}\gamma^{-1}=\left[\begin{array}{cc}
0 & -iq_{0}\\
iq_{0} & 0\end{array}\right]=\left[\begin{array}{cc}
0 & -i\\
i & 0\end{array}\right]\qquad\gamma_{\pi}\gamma^{-1}=\left[\begin{array}{cc}
0 & -q_{\pi}\\
-q_{\pi} & 0\end{array}\right]\]
 As shown above, this Clifford algebra is also faithfully represented
by $2\times2$ matrices of quaternions. All the odd graded basis elements
are non-zero only in off-diagonal blocks, and all even graded elements
are non-zero only in diagonal blocks. This is the sense in which a
representation is {}``chiral'' -- a useful property. An arbitrary
$Cl_{1,3}$ element can be written as\begin{align}
C & =C_{s}+C_{v}^{0}\gamma_{0}+C_{v}^{\pi}\gamma_{\pi}+C_{b}^{0\pi}\gamma_{0\pi}+\frac{1}{2}C_{b}^{\pi\rho}\gamma_{\pi\rho}+C_{t}^{0}\gamma_{0}\gamma^{-1}+C_{t}^{\pi}\gamma_{\pi}\gamma^{-1}+C_{p}\gamma\nonumber \\
 & =\left[\begin{array}{cc}
\left(C_{s}-iC_{p}\right)q_{0}+\left(iC_{b}^{0\sigma}-\frac{1}{2}C_{b}^{\pi\rho}\epsilon_{\pi\rho}{}^{\sigma}\right)q_{\sigma} & \left(C_{v}^{0}-iC_{t}^{0}\right)q_{0}+\left(-iC_{v}^{\pi}-C_{t}^{\pi}\right)q_{\pi}\\
\left(C_{v}^{0}+iC_{t}^{0}\right)q_{0}+\left(iC_{v}^{\pi}-C_{t}^{\pi}\right)q_{\pi} & \left(C_{s}+iC_{p}\right)q_{0}+\left(-iC_{b}^{0\sigma}-\frac{1}{2}C_{b}^{\pi\rho}\epsilon_{\pi\rho}{}^{\sigma}\right)q_{\sigma}\end{array}\right]\nonumber \\
 & =\left[\begin{array}{cc}
C_{e}^{0}q_{0}+C_{e}^{\sigma}q_{\sigma} & C_{o}^{0}q_{0}+C_{o}^{\sigma}q_{\sigma}\\
C_{o}^{0*}q_{0}+C_{o}^{\sigma*}q_{\sigma} & C_{e}^{0*}q_{0}+C_{e}^{\sigma*}q_{\sigma}\end{array}\right]=\left[\begin{array}{cc}
C_{Te}+C_{Se} & C_{To}+C_{So}\\
C_{To}^{\dagger}-C_{So}^{\dagger} & C_{Te}^{\dagger}-C_{Se}^{\dagger}\end{array}\right]\nonumber \\
 & =\left[\begin{array}{cc}
C_{e}^{\mu}q_{\mu} & C_{o}^{\mu}q_{\mu}\\
C_{o}^{\mu*}q_{\mu} & C_{e}^{\mu*}q_{\mu}\end{array}\right]=\left[\begin{array}{cc}
C_{e}^{L} & C_{o}^{R}\\
C_{o}^{L} & C_{e}^{R}\end{array}\right]\label{eq:matdeco}\end{align}
 a $2\times2$ matrix of quaternions with complex coefficients. The
$e/o$ labels stand for even and odd Clifford grade components, the
$T/S$ labels for time and space components, and the $L/R$ labels
for left and right chiralities. Each matrix element consists of complex
coefficients multiplying quaternions. If charge conjugation is defined
using the very useful rule,\begin{equation}
-q_{2}q_{\mu}^{*}q_{2}=q_{\mu}\label{eq:quatconj}\end{equation}
 such that it acts as complex conjugation on the coefficients, but
not within the matrices representing the quaternions, the relationship
between left and right components of any Clifford element may be written
as\begin{align}
\overline{C_{e}^{L}} & =\overline{\left(C_{e}^{\mu}q_{\mu}\right)}=-q_{2}\left(C_{e}^{L}\right)^{*}q_{2}=C_{e}^{\mu*}q_{\mu}=C_{e}^{R}\nonumber \\
\overline{C_{o}^{R}} & =\overline{\left(C_{o}^{\mu}q_{\mu}\right)}=-q_{2}\left(C_{o}^{R}\right)^{*}q_{2}=C_{o}^{\mu*}q_{\mu}=C_{o}^{L}\label{eq:chiralpart}\end{align}
 The 16 real variables in an element, $C$, of $Cl_{1,3}$ are thus
encoded by the 4 complex variables each in $C_{e}^{L}$ and $C_{o}^{R}$.
Also, since $C_{e}^{R}=\overline{C_{e}^{L}},$ the scalar part of
a $4\times4$ Clifford element is equal to the scalar part of the
associated complex quaternionic ($2\times2$) even representative.\begin{equation}
\left\langle C\right\rangle =C_{s}=\left\langle C_{e}^{L}\right\rangle \label{eq:chirals}\end{equation}
 since the scalar part operator also returns just the real part.

If we were to work in $Cl_{4,0}$, obtained by changing the vector
basis representatives, $\gamma_{\pi}\rightarrow i\gamma_{\pi}$, the
resulting $2\times2$ chiral matrix would consist of quaternions with
real coefficients. There is a representation for $Cl_{1,3}$ of quaternions
with real coefficients, but it's not chiral -- the even and odd graded
components mix between diagonal and off-diagonal blocks. Nevertheless,
some Clifford algebras decompose into a quaternionic {}``S-sub-algebra''
times a Clifford algebra of lower grade, such as the case with $Cl_{1,3}=\mathbb{H}\otimes Cl_{2,0}$.
The basis vectors for this $Cl_{2,0}$ corresponding to the Weyl representation
is found from\[
\gamma_{0}=q_{0}\otimes\left[\begin{array}{cc}
0 & 1\\
1 & 0\end{array}\right]\qquad\gamma_{\pi}=q_{\pi}\otimes\left[\begin{array}{cc}
0 & -i\\
i & 0\end{array}\right]\]
 Although the choice of representation is a useful device for discerning
structure, all physical Lagrangians are invariant under the global
version of the Clifford adjoint (\ref{eq:adj}), and therefore under
a change of representation for the basis matrices.

Clifford algebra projectors may be built by combining elements having
a diagonal representation, such as the chiral projector, $P^{L/R}=\frac{1}{2}\left(1\pm i\gamma\right)$,
that projects out collections of {}``left-acting'' elements, \[
CP^{L}=\left[\begin{array}{cc}
C_{e}^{L} & C_{o}^{R}\\
C_{o}^{L} & C_{e}^{R}\end{array}\right]\left[\begin{array}{cc}
{\textbf{1}} & 0\\
0 & 0\end{array}\right]=\left[\begin{array}{cc}
C_{e}^{L} & 0\\
C_{o}^{L} & 0\end{array}\right]\]
 The explicit appearance of {}``$i$'' in constructing some projectors
implies the use of the corresponding complex Clifford algebra -- Clifford
algebras with complex coefficients. Projectors may also be used to
create mixed basis elements, useful for breaking elements up into
two parts such as\begin{align*}
C_{b} & =\left[\begin{array}{cc}
C_{b}^{\sigma}q_{\sigma}\\
 & C_{b}^{\sigma*}q_{\sigma}\end{array}\right]=\left(C_{b}^{0\pi}\gamma_{0\pi}+\frac{1}{2}C_{b}^{\pi\rho}\gamma_{\pi\rho}\right)\left(P^{L}+P^{R}\right)\\
 & =\left(iC_{b}^{0\sigma}-\frac{1}{2}C_{b}^{\pi\rho}\epsilon_{\pi\rho}{}^{\sigma}\right)\left[\begin{array}{cc}
q_{\sigma}\\
 & 0\end{array}\right]+\left(-iC_{b}^{0\sigma}-\frac{1}{2}C_{b}^{\pi\rho}\epsilon_{\pi\rho}{}^{\sigma}\right)\left[\begin{array}{cc}
0\\
 & q_{\sigma}\end{array}\right]\\
 & =C_{b}^{\sigma}\frac{1}{2}\left(-i\gamma_{0\sigma}-\epsilon^{\pi\rho}{}_{\sigma}\gamma_{\pi\rho}\right)+C_{b}^{\sigma*}\left(i\gamma_{0\sigma}-\epsilon^{\pi\rho}{}_{\sigma}\gamma_{\pi\rho}\right)\\
 & =C_{b}^{\sigma}T_{\sigma}^{L}+C_{b}^{\sigma*}T_{\sigma}^{R}\end{align*}
 a bivector broken up into its left-chiral (self-dual) and right-chiral
(anti-self-dual) parts, with chiral ((anti-)self-dual) basis elements
satisfying $T_{\sigma}^{L/R}P^{L/R}=T_{\sigma}^{L/R}=\pm T_{\sigma}^{L/R}i\gamma^{-1}$.
Although it is possible to equate $C_{b}^{\sigma}T_{\sigma}^{L}$
and $C_{b}^{L}=C_{b}^{\sigma}q_{\sigma}$ in most computations, care
should be taken since $q_{\sigma}$ is a $2\times2$ matrix and $T_{\sigma}^{L}$
a $4\times4$ matrix with non-zero elements equal to $q_{\sigma}$.

\subsection{Chiral Gravity}

The Clifford connection for $Cl_{1,3}$ in the Weyl rep is\[
\f{\Omega}=\f{e}+\f{\omega}=\left[\begin{array}{cc}
\f{\omega^{L}} & \f{e^{R}}\\
\f{e^{L}} & \f{\omega^{R}}\end{array}\right]\]
 with quaternionic elements times complex coefficients, equivalent
to $2\times2$ complex matrices, equal to\begin{align*}
\f{e^{R}} & =\f{e^{0}}q_{0}-i\f{e^{\pi}}q_{\pi}\\
\f{\omega^{L}} & =\left(i\f{\omega^{0\sigma}}-\frac{1}{2}\f{\omega^{\pi\rho}}\epsilon_{\pi\rho}{}^{\sigma}\right)q_{\sigma}\end{align*}
 and their chiral partners given by (\ref{eq:chiralpart}). The curvature
is\[
\f{\f{F}}=\f{\partial}\f{\Omega}+\frac{1}{2}\f{\Omega}\f{\Omega}=\f{\f{T}}+\f{\f{R}}+\f{\f{E}}=\left[\begin{array}{cc}
\f{\f{R^{L}}}+\f{\f{E^{L}}} & \f{\f{T^{R}}}\\
\f{\f{T^{L}}} & \f{\f{R^{R}}}+\f{\f{E^{R}}}\end{array}\right]\]
 with\begin{align*}
\f{\f{T^{R}}} & =\f{\partial}\f{e^{R}}+\frac{1}{2}\f{\omega^{L}}\f{e^{R}}+\frac{1}{2}\f{e^{R}}\f{\omega^{R}}\\
\f{\f{E^{L}}} & =\frac{1}{2}\f{e^{R}}\f{e^{L}}=\frac{1}{2}\left(\f{e^{0}}q_{0}-i\f{e^{\pi}}q_{\pi}\right)\left(\f{e^{0}}q_{0}+i\f{e^{\rho}}q_{\rho}\right)=\left(i\f{e^{0}}\f{e^{\sigma}}+\f{e^{\pi}}\f{e^{\rho}}\epsilon_{\pi\rho}{}^{\sigma}\right)q_{\sigma}\\
\f{\f{R^{L}}} & =\f{\partial}\f{\omega^{L}}+\frac{1}{2}\f{\omega^{L}}\f{\omega^{L}}\end{align*}
 The gravitational Lagrangian (\ref{eq:ffg}), using (\ref{eq:chirals})
and (\ref{eq:pseudo}), is\begin{align*}
\ub{L} & =\frac{1}{2}\left\langle \f{\f{F}}\f{\f{F}}\gamma^{-1}\right\rangle =\frac{1}{2}\left\langle i\left(\f{\f{R^{L}}}+\f{\f{E^{L}}}\right)\left(\f{\f{R^{L}}}+\f{\f{E^{L}}}\right)+i\f{\f{T^{R}}}\overline{\f{\f{T^{R}}}}\right\rangle \\
 & =\left\langle i\f{\f{E^{L}}}\f{\f{R^{L}}}+\frac{i}{2}\f{\f{E^{L}}}\f{\f{E^{L}}}\right\rangle +\f{\partial}\ub{b}\\
 & =\left\langle i\f{\f{E^{L}}}\f{\partial}\f{\omega^{L}}-\left(-\frac{i}{2}\f{\f{E^{L}}}\f{\omega^{L}}\f{\omega^{L}}-\frac{i}{2}\f{\f{E^{L}}}\f{\f{E^{L}}}\right)\right\rangle +\f{\partial}\ub{b}\end{align*}
 the chiral (also known as {}``self dual'') Lagrangian for gravity
plus Chern-Simons boundary term. The new dynamical variables, $i\f{\f{E^{L}}}$
and $\f{\omega^{L}}$, are spatial quaternions with
complex coefficients encoding the same information as $\f{e}$
and $\f{\omega}$. The Ashtekar Hamiltonian for gravity
with cosmological term, appearing above, is\[
\ub{H}=\left\langle -\frac{1}{2}i\f{\f{E^{L}}}\f{\omega^{L}}\f{\omega^{L}}+\frac{i}{2}i\f{\f{E^{L}}}i\f{\f{E^{L}}}\right\rangle \]
 with resulting canonical equations of motion,

\begin{align*}
\f{\partial}\f{\omega^{L}} & =\frac{\partial}{\partial i\f{\f{E^{L}}}}\ub{H}=-\frac{1}{2}\f{\omega^{L}}\f{\omega^{L}}-\f{\f{E^{L}}}\\
\f{\partial}i\f{\f{E^{L}}} & =\frac{\partial}{\partial\f{\omega^{L}}}\ub{H}=-\f{\omega^{L}}\times i\f{\f{E^{L}}}\end{align*}

\subsection{Spinors}

A spinor in four dimensions is conventionally defined as a column
of four complex Grassmann valued (anti-commuting) numbers, which in
the chiral representation breaks into two Weyl column spinors,\[
\psi^{|}=\left[\begin{array}{c}
\psi_{L\uparrow}\\
\psi_{L\downarrow}\\
\psi_{R\uparrow}\\
\psi_{R\downarrow}\end{array}\right]=\left[\begin{array}{c}
\psi_{L}\\
\psi_{R}\end{array}\right]\]
 The real valued Dirac Lagrangian in curved spacetime naturally splits
and mixes chiral components,\begin{align}
\ub{L}{}_{\textrm{Dirac}} & =\ub{e}\left\langle \left(\psi^{|}\right)^{\dagger}\gamma_{0}\vg{e}\left(\f{\partial}+\frac{1}{2}\f{\omega}-i\f{A}+\frac{im}{4}\f{e}\right)\psi^{|}\right\rangle \label{eq:diracwe}\\
 & =\ub{e}\left\langle \left[\begin{array}{cc}
\left(\psi_{L}\right)^{\dagger} & \left(\psi_{R}\right)^{\dagger}\end{array}\right]\left[\begin{array}{cc}
 & 1\\
1\end{array}\right]\left[\begin{array}{cc}
 & \vg{e}_{R}\\
\vg{e}_{L}\end{array}\right]\left[\begin{array}{cc}
\f{\partial}+\frac{1}{2}\f{\omega^{L}}-i\f{A} & \frac{im}{4}\f{e^{R}}\\
\frac{im}{4}\f{e^{L}} & \f{\partial}+\frac{1}{2}\f{\omega^{R}}-i\f{A}\end{array}\right]\left[\begin{array}{c}
\psi_{L}\\
\psi_{R}\end{array}\right]\right\rangle \nonumber \\
 & =\ub{e}\left\langle \left(\psi_{L}\right)^{\dagger}\left[\vg{e}_{L}\left(\f{\partial}+\frac{1}{2}\f{\omega^{L}}-i\f{A}\right)\psi_{L}+im\psi_{R}\right]+\left(\psi_{R}\right)^{\dagger}\left[\vg{e}_{R}\left(\f{\partial}+\frac{1}{2}\f{\omega^{R}}-i\f{A}\right)\psi_{R}+im\psi_{L}\right]\right\rangle \nonumber \end{align}
 Note the novel appearance of the mass term in the Dirac operator,
(\ref{eq:diracwe}), consistent with $\f{\Omega}=\f{\omega}+\frac{im}{2}\f{e}$,
as well as a $u(1)$ gauge field with generator $-i$. The Dirac Lagrangian
is invariant under charge conjugation, $\psi^{|}\rightarrow\psi^{|c}=i\gamma_{2}(\psi^{|})^{*}$,
with\begin{align*}
\psi_{L}^{c} & =q_{2}\psi_{R}^{*}=\overline{\psi_{R}}\\
\psi_{R}^{c} & =-q_{2}\psi_{L}^{*}=\overline{\psi_{L}}\\
\f{\left(-iA\right)^{c}} & =\f{\left(-iA\right)^{*}}\end{align*}
 The anti-particle partner to any two-component fermion is labeled
by an overline, similar to the labeling for quaternions. This invariance
can be confirmed using (\ref{eq:quatconj}) to get\begin{align*}
-q_{2}\vg{e}_{L}^{*}q_{2} & =\overline{\vg{e}_{L}}=\vg{e}_{R}\\
-q_{2}\f{\omega^{L*}}q_{2} & =\overline{\f{\omega^{L}}}=\f{\omega^{R}}\end{align*}
 and then{\footnotesize \begin{align*}
\ub{L}{}'_{\textrm{Dirac}} & =\ub{e}\left\langle \left(\psi_{L}^{c}\right)^{\dagger}\left[\vg{e}_{L}\left(\f{\partial}+\frac{1}{2}\f{\omega^{L}}+\f{\left(-iA\right)^{c}}\right)\psi_{L}^{c}+im\psi_{R}^{c}\right]+\left(\psi_{R}^{c}\right)^{\dagger}\left[\vg{e}_{R}\left(\f{\partial}+\frac{1}{2}\f{\omega^{R}}+\f{\left(-iA\right)^{c}}\right)\psi_{R}^{c}+im\psi_{L}^{c}\right]\right\rangle \\
 & =\ub{e}\left\langle -\left(\psi_{R}^{*}\right)^{\dagger}q_{2}\left[\vg{e}_{L}\left(\f{\partial}+\frac{1}{2}\f{\omega^{L}}+i\f{A^{*}}\right)q_{2}\psi_{R}^{*}-imq_{2}\psi_{L}^{*}\right]+\left(\psi_{L}^{*}\right)^{\dagger}q_{2}\left[-\vg{e}_{R}\left(\f{\partial}+\frac{1}{2}\f{\omega^{R}}+i\f{A^{*}}\right)q_{2}\psi_{L}^{*}+imq_{2}\psi_{R}^{*}\right]\right\rangle \\
 & =\ub{e}\left\langle -\left(\psi_{R}\right)^{\dagger}q_{2}\left[\vg{e}_{L}^{*}\left(\f{\partial}+\frac{1}{2}\f{\omega^{L*}}-i\f{A}\right)q_{2}\psi_{R}+imq_{2}\psi_{L}\right]+\left(\psi_{L}\right)^{\dagger}q_{2}\left[-\vg{e}_{R}^{*}\left(\f{\partial}+\frac{1}{2}\f{\omega^{R*}}-i\f{A}\right)q_{2}\psi_{L}-imq_{2}\psi_{R}\right]\right\rangle \\
 & =\ub{e}\left\langle \left(\psi_{R}\right)^{\dagger}\left[\vg{e}_{R}\left(\f{\partial}+\frac{1}{2}\f{\omega^{L}}-i\f{A}\right)\psi_{R}+im\psi_{L}\right]+\left(\psi_{L}\right)^{\dagger}\left[\vg{e}_{L}\left(\f{\partial}+\frac{1}{2}\f{\omega^{R}}-i\f{A}\right)\psi_{L}+im\psi_{R}\right]\right\rangle =\ub{L}{}{}_{\textrm{Dirac}}\end{align*}
}Note in the above that the anti-partner to a left chiral fermion
is a right chiral fermion, and vice versa.

The Dirac Lagrangian is also invariant under chirality conjugation,
$L\leftrightarrow R$. However, the weak force breaks this invariance
and it is useful to construct rectangular blocks of spinors to describe
its interaction. Conventionally, these blocks are single columns of
two component spinors (Clifford ideals); however, it is also natural
to represent multiple fermions using more than one column in a single
Clifford field. For example, a coupling between a left chiral gauge
field, $\f{W^{L}}=-i\f{W^{\pi}}\Sigma_{\pi}$,
and a two element high block of two component spinors is conventionally
written as{\footnotesize \begin{equation}
\ub{L}{}_{W\psi}=\ub{e}\left\langle \left(\psi_{L}\right)^{\dagger}\vg{e}_{L}\frac{1}{2}\f{W^{L}}\psi_{L}\right\rangle =\ub{e}\left\langle \left[\begin{array}{cc}
\nu_{L}^{e\dagger} & e_{L}^{\dagger}\end{array}\right]\left[\begin{array}{cc}
-\frac{i}{2}\vg{e}_{L}\f{W^{3}} & -\frac{i}{2}\vg{e}_{L}\left(\f{W^{1}}-i\f{W^{2}}\right)\\
-\frac{i}{2}\vg{e}_{L}\left(\f{W^{1}}+i\f{W^{2}}\right) & \frac{i}{2}\vg{e}_{L}\f{W^{3}}\end{array}\right]\left[\begin{array}{c}
\nu_{L}^{e}\\
e_{L}\end{array}\right]\right\rangle \label{eq:wp}\end{equation}
}with $\Sigma_{\pi}$ equal to a $4\times4$ block representation
of the Pauli matrices. It is natural in the Clifford algebra context
to replace the $4\times1$ block of fermion components with a square
$4\times4$ block containing new particles as well as appropriate
anti-particle partners, such as\[
\psi_{L}=\left[\begin{array}{cc}
\left[\begin{array}{cc}
\nu_{L}^{e} & -\overline{e_{R}}\end{array}\right] & \left[\begin{array}{cc}
u_{L} & -\overline{d_{R}}\end{array}\right]\\
\left[\begin{array}{cc}
e_{L} & \overline{v_{R}^{e}}\end{array}\right] & \left[\begin{array}{cc}
d_{L} & \overline{u_{R}}\end{array}\right]\end{array}\right]\]
 This gives the new particles interacting in the same way with the
gauge field in (\ref{eq:wp}). Each $2\times2$ quadrant of this block
of fermions may be interpreted as a quaternion with complex coefficients.

\section{BRST gauge fixing}

The BRST method fixes and accounts for gauge symmetries by introducing
new fields, with anti-commuting (Grassmann valued) coefficients, having
dynamics and interactions with existing fields that breaks the original
local gauge symmetry but includes a new global (super) symmetry involving
a {}``rotation'' between old and new fields. This method of gauge
fixing is an indispensable tool in the application of path integral
methods in the quantum field theory of non-abelian gauge fields, and
has a natural extension to describe the existence and dynamics of
fermionic spinor fields \cite{lisi}.

A restricted BF Lagrangian,\[
\ub{L}=\left\langle \f{\f{B}}\left(\f{\f{F}}-\f{\f{\Phi}}(A,B)\right)\right\rangle \]
 which is invariant off shell, $\delta_{C^{o}}\ub{L}=0$, under
some subset of the gauge transformation (\ref{eq:gtransf}), such
as the subset formed by odd graded gauge parameter fields, $C=C^{o}$,
is amenable to the BRST method. Note that the gravitational Lagrangian
does not satisfy this condition for odd $C^{o}$, so it is a more
general Lagrangian, involving a higher dimensional Clifford algebra,
being considered here. The BRST method proceeds by introducing a {}``ghost''
field, $C_{g}^{o}$, with Grassmann valued (anti-commuting) coefficients
but otherwise equivalent to $C^{o}$, and a Grassmann valued conjugate
3-form field, $\ub{B}{}_{g}^{o}$, as well as a real valued
partner field, $\ub{\lambda}{}^{o}$. The new system is equipped
with a global BRST transformation, a {}``super-symmetry rotation''
between real and Grassmann valued variables,\begin{align*}
\delta_{g}\f{A} & =-\f{\nabla}C_{g}^{o} & \delta_{g}C_{g}^{o} & =-\frac{1}{2}C_{g}^{o}\times C_{g}^{o}\\
\delta_{g}\f{\f{B}} & =C_{g}^{o}\times\f{\f{B}} & \delta_{g}\ub{B}{}_{g}^{o} & =-i\ub{\lambda}{}^{o}\\
\delta_{g}\ub{\lambda}{}^{o} & =0\end{align*}
 nilpotent by design, $\delta_{g}\delta_{g}=0$, and leaving the Lagrangian
invariant. Dynamics are introduced for the ghosts by adding a {}``BRST
exact'' term to get a BRST extended Lagrangian,\[
\ub{L'}=\ub{L}+\delta_{g}\ub{\Psi}\]
 with some BRST potential chosen, such as\[
\ub{\Psi}=\left\langle i\ub{B}{}_{g}^{o}\f{A}\right\rangle \]
 which gives\[
\delta_{g}\ub{\Psi}=\left\langle \ub{\lambda}{}^{o}\f{A}\right\rangle +\left\langle i\ub{B}{}_{g}^{o}\f{\nabla}C_{g}^{o}\right\rangle \]
 The BRST partner variable, $\ub{\lambda}{}^{o}$, acts as a
Lagrange multiplier constraining the connection to be even, $\f{A}=\f{A^{e}}$
-- fixing the gauge freedom. The resulting effective Lagrangian is\[
\ub{L^{\textrm{eff}}}=\left\langle \f{\f{B^{e}}}\left(\f{\f{F^{e}}}-\f{\f{\Phi}}(A^{e},B^{e})\right)\right\rangle +\left\langle i\ub{B}{}_{g}^{o}\f{\nabla^{e}}C_{g}^{o}\right\rangle \]
 The new fields and this Lagrangian are compatible with a Poisson
bracket modified to include the canonical pair of ghost fields and
the Hamiltonian\[
\ub{H^{\textrm{eff}}}=\ub{H^{e}}-\left\langle i\ub{B}{}_{g}^{o}\left(\f{A^{e}}\times C_{g}^{o}\right)\right\rangle \]
 as well as a generator for the BRST transformation. The structure
of this Lagrangian suggests the construction of a BRST restricted
and extended connection ({}``super connection''), $\f{\tilde{A}}=\f{A^{e}}+C_{g}^{o}$,
having curvature\[
\f{\f{\tilde{F}}}=\f{\partial}\f{\tilde{A}}+\frac{1}{2}\f{\tilde{A}}\times\f{\tilde{A}}=\left(\f{\partial}\f{A^{e}}+\frac{1}{2}\f{A^{e}}\times\f{A^{e}}\right)+\left(\f{\partial}C_{g}^{o}+\f{A^{e}}\times C_{g}^{o}\right)+\frac{1}{2}C_{g}^{o}\times C_{g}^{o}=\f{\f{F^{e}}}+\f{\nabla^{e}}C_{g}^{o}+\frac{1}{2}C_{g}^{o}\times C_{g}^{o}\]
 and giving the extended BF Lagrangian as\begin{equation}
\ub{L^{\textrm{eff}}}=\left\langle \left(\f{\f{B^{e}}}+i\ub{B}{}_{g}^{o}\right)\left(\f{\f{F^{e}}}-\f{\f{\Phi}}(A^{e},B^{e})+\f{\nabla^{e}}C_{g}^{o}\right)\right\rangle =\left\langle \f{\f{\tilde{B}}}\left(\f{\f{\tilde{F}}}-\f{\f{\tilde{\Phi}}}(A^{e},B^{e},C_{g}^{o})\right)\right\rangle =\left\langle \f{\f{\tilde{B}}}\f{\partial}\f{\tilde{A}}\right\rangle -\ub{H^{\textrm{eff}}}\label{eq:bfext}\end{equation}

With a {}``chiral'' Clifford algebra representation, splitting into
even and odd quadrants, the extended connection can be written in
blocks as\begin{equation}
\tilde{\f{A}}=\f{A^{e}}+C_{g}^{o}=\left[\begin{array}{cc}
\f{A^{l}} & C_{g}^{r}\\
C_{g}^{l} & \f{A^{r}}\end{array}\right]\label{eq:extendedA}\end{equation}
 with curvature\begin{equation}
\f{\f{\tilde{F}}}=\left[\begin{array}{cc}
\f{\f{F^{l}}}+\frac{1}{2}C_{g}^{r}\times C_{g}^{l} & \f{\partial}C_{g}^{r}+\frac{1}{2}\f{A^{l}}C_{g}^{r}-\frac{1}{2}C_{g}^{r}\f{A^{r}}\\
\f{\partial}C_{g}^{l}+\frac{1}{2}\f{A^{r}}C_{g}^{l}-\frac{1}{2}C_{g}^{l}\f{A^{l}} & \f{\f{F^{r}}}+\frac{1}{2}C_{g}^{l}\times C_{g}^{r}\end{array}\right]\label{eq:extF}\end{equation}
 We may re-label the variable $C_{g}^{r}=\psi$ and interpret this
field as a block of fermionic spinor fields, and write the conjugate
BRST field as $\ub{B}{}_{g}^{l}=-i\ub{e}\chi\gamma_{0}\vg{e_{l}}$,
in terms of another block of spinor fields, $\chi$. Presuming that
$\f{e}$ is independent of the BRST transformation,
the fermionic Lagrangian term is proportional to\[
\ub{L^{\textrm{f}}}=\left\langle i\ub{B}{}_{g}^{o}\f{\nabla^{e}}C_{g}^{o}\right\rangle \sim\ub{e}\left\langle \chi\gamma_{0}\vg{e_{l}}\left(\f{\partial}\psi+\frac{1}{2}\f{A^{l}}\psi-\frac{1}{2}\psi\f{A^{r}}\right)\right\rangle \]
 with one block of gauge fields operating on the fermion block from
the left, the {}``left-acting'' $\f{A^{l}}$ , and
one from the right, the {}``right-acting'' $\f{A^{r}}$.
In this manner the BRST method produces blocks of odd graded fermions
with left and right acting blocks of gauge fields.

\section{The standard model and gravity}

The standard model of gauge forces and fermions is constructed by
considering Clifford algebra fibers of higher dimension, still over
a four dimensional base manifold. One Clifford algebra in particular,
$Cl_{1,7}=\mathbb{H}\otimes Cl_{0,6}$, has a quaternionic decomposition
as well as an appealing bivector sub-algebra, $spin(6)=su(4)$. Applying
the methods and tools established so far leads to a concise description
of the standard model compatible with the elegant description put
forth by Greg Trayling \cite{trayling}, with a few cosmetic modifications
and the natural inclusion of gravity.

\subsection{Basis vectors}

The $Cl_{1,7}$ algebra has representations as real $16\times16$
matrices, restricted complex $16\times16$ matrices, $8\times8$ matrices
of quaternions, or as restricted $8\times8$ matrices of quaternions
with complex coefficients -- similar to (\ref{eq:weylrep}). Inspired
by the Weyl representation, the eight Hermitian and anti-Hermitian
Clifford basis vectors, $\Gamma_{\alpha}$, are chosen to be{\footnotesize \[
\Gamma_{0}+\Gamma_{\pi}+z\Gamma_{4}+a\Gamma_{5}+b\Gamma_{6}+c\Gamma_{7}=\left[\begin{array}{cccccccc}
 &  &  &  & 1-iq_{\pi} &  & -c+iz & -a+ib\\
 &  &  &  &  & 1-iq_{\pi} & -a-ib & c+iz\\
 &  &  &  & -c-iz & -a+ib & 1+iq_{\pi}\\
 &  &  &  & -a-ib & c-iz &  & 1+iq_{\pi}\\
1+iq_{\pi} &  & c-iz & a-ib\\
 & 1+iq_{\pi} & a+ib & -c-iz\\
c+iz & a-ib & 1-iq_{\pi}\\
a+ib & -c+iz &  & 1-iq_{\pi}\end{array}\right]\]
}in which the higher Clifford basis vector elements, $\Gamma_{\phi=4,5,6,7}$,
are directly related to the $4\times4$ block Pauli sigma matrices,
$\Sigma_{\phi-4}$, similar to the way the lower basis vectors, $\Gamma_{\mu=0,1,2,3}$,
are related to the corresponding $2\times2$ Pauli matrices or quaternions.
The resulting pseudo-scalar is\[
\Gamma=\Gamma_{0}\Gamma_{1}\Gamma_{2}\Gamma_{3}\Gamma_{4}\Gamma_{5}\Gamma_{6}\Gamma_{7}=\left[\begin{array}{cc}
-i\\
 & i\end{array}\right]\]
 used to build the fundamental left/right projector, $P^{l/r}=\frac{1}{2}\left(1\pm i\Gamma\right)$.
This projector necessarily contains $i$, and its use implies consideration
of the corresponding complex Clifford algebra, $\mathbb{C}l_{8}$.
The Left/Right {}``chiral'' projector, $P^{L/R}=\frac{1}{2}\left(1\pm\gamma_{5}\right)$,
comes from\[
\gamma_{5}=\Gamma_{4}\Gamma_{5}\Gamma_{6}\Gamma_{7}=i\gamma=-\Gamma_{0}\Gamma_{1}\Gamma_{2}\Gamma_{3}\Gamma=\left[\begin{array}{cccc}
1\\
 & -1\\
 &  & 1\\
 &  &  & -1\end{array}\right]\]

\subsection{Trayling's model plus gravity}

Following Trayling, a set of Clifford basis elements are chosen that
reproduce the familiar anti-Hermitian $su(2)_{L/R}$ and $su(3)$
generators as well as those for gravity. Remarkably, the necessary
generators can all be constructed from the 28 bivectors of $Cl_{1,7}$
projected into left and right acting blocks via $P^{l/r}$ -- but
therefore implying use of $\mathbb{C}l_{8}$. Almost all of the 28
bivectors are used to construct the left acting generators, while
a subgroup of 8 right acting generators are picked out corresponding
to the complex conjugates of the eight desired Gell-Mann matrices,
as well as a final generator having non-zero elements in both left
and right blocks:\begin{equation}
\begin{array}{rclcrcl}
T_{\mu\phi} & = & \Gamma_{\mu\phi}P^{l} &  & T'_{1} & = & \frac{1}{2}_{{}}\left(\Gamma_{16}-\Gamma_{25}\right)P^{r}\\
T_{\mu\nu} & = & \Gamma_{\mu\nu}P^{l} &  & T'_{2} & = & \frac{1}{2}_{{}}\left(-\Gamma_{15}-\Gamma_{26}\right)P^{r}\\
T_{1}^{L} & = & \frac{1}{2}\left(-\Gamma_{45}+\Gamma_{67}\right)P^{l} &  & T'_{3} & = & \frac{1}{2}_{{}}\left(\Gamma_{12}-\Gamma_{56}\right)P^{r}\\
T_{2}^{L} & = & \frac{1}{2}\left(-\Gamma_{46}-\Gamma_{57}\right)P^{l} &  & T'_{4} & = & \frac{1}{2}_{{}}\left(-\Gamma_{14}+\Gamma_{27}\right)P^{r}\\
T_{3}^{L} & = & \frac{1}{2}\left(-\Gamma_{47}+\Gamma_{56}\right)P^{l} &  & T'_{5} & = & \frac{1}{2}_{{}}\left(\Gamma_{17}+\Gamma_{24}\right)P^{r}\\
T_{1}^{R} & = & \frac{1}{2}\left(\Gamma_{45}+\Gamma_{67}\right)P^{l} &  & T'_{6} & = & \frac{1}{2}_{{}}\left(\Gamma_{45}+\Gamma_{67}\right)P^{r}\\
T_{2}^{R} & = & \frac{1}{2}\left(\Gamma_{46}-\Gamma_{57}\right)P^{l} &  & T'_{7} & = & \frac{1}{2}_{{}}\left(-\Gamma_{46}+\Gamma_{57}\right)P^{r}\\
T_{0} & = & \frac{1}{2}\left(\Gamma_{47}+\Gamma_{56}\right)P^{l}+\frac{1}{3}\left(\Gamma_{12}-\Gamma_{47}+\Gamma_{56}\right)P^{r} & \qquad & T'_{8} & = & \frac{1}{2\sqrt{3}}\left(\Gamma_{12}+2\Gamma_{47}+\Gamma_{56}\right)P^{r}\end{array}\label{eq:gen8}\end{equation}
 The action for the standard model is presumed to allow odd graded
connection elements to be supplanted by the use of an odd graded BRST
field (\ref{eq:extendedA}). The resulting BRST extended connection,
built of selected projected bivector generators from (\ref{eq:gen8})
and an odd graded block of fermions, with one quadrant left undetermined
because it's not clear what should go in it, takes the form\begin{align}
\tilde{\f{A}} & =\phi\f{e}+\f{\omega}+\f{W}+\f{B}+\f{G}+\psi\label{eq:matrix}\\
 & =\phi^{\psi}\f{e^{\mu}}T_{\mu\psi}+\frac{1}{2}\f{\omega^{\mu\nu}}T_{\mu\nu}+\f{W^{\pi}}T_{\pi}^{L}+\f{B}T_{0}+\f{G^{A}}T'_{A}+\nu^{e}+e+u+d+\overline{\nu^{e}}+\overline{e}+\overline{u}+\overline{d}\qquad\qquad\qquad\qquad \: \nonumber \end{align}
{\footnotesize \[
\quad \: = \left[\begin{array}{cccccccc}
\f{\omega^{L}}-i\f{W^{3}} & -i\f{W^{1}}-\f{W^{2}} & \phi_{2}^{0}\f{e^{R}} & \phi_{1}^{+}\f{e^{R}} & \nu_{L}^{e} & u_{L}^{r} & u_{L}^{g} & u_{L}^{b}\\
-i\f{W^{1}}+\f{W^{2}} & \f{\omega^{L}}+i\f{W^{3}} & \phi_{2}^{+}\f{e^{R}} & \phi_{1}^{0}\f{e^{R}} & e_{L} & d_{L}^{r} & d_{L}^{g} & d_{L}^{b}\\
-\phi_{1}^{0}\f{e^{L}} & \phi_{1}^{+}\f{e^{L}} & \f{\omega^{R}}-i\f{B} &  & \nu_{R}^{e} & u_{R}^{r} & u_{R}^{g} & u_{R}^{b}\\
\phi_{2}^{+}\f{e^{L}} & -\phi_{2}^{0}\f{e^{L}} &  & \f{\omega^{R}}+i\f{B} & e_{R} & d_{R}^{r} & d_{R}^{g} & d_{R}^{b}\\
 &  &  &  & -i\f{B}\\
 &  &  &  &  & \frac{i}{3}\f{B}-i\f{G^{3}}-\frac{i}{\sqrt{3}}\f{G^{8}} & -i\f{G^{1}}+\f{G^{2}} & -i\f{G^{4}}+\f{G^{5}}\\
 &  &  &  &  & -i\f{G^{1}}-\f{G^{2}} & \frac{i}{3}\f{B}+i\f{G^{3}}-\frac{i}{\sqrt{3}}\f{G^{8}} & -iG^{6}+\f{G^{7}}\\
 &  &  &  &  & -i\f{G^{4}}-\f{G^{5}} & -i\f{G^{6}}-\f{G^{7}} & \frac{i}{3}\f{B}+\frac{2i}{\sqrt{3}}\f{G^{8}}\end{array}\right]\]
}This is the extended connection for the standard model and gravity.
Each $2\times2$ fermionic block (a complex quaternion) includes an
anti-particle that isn't shown, such as\[
e_{L}\leftrightarrow\left[\begin{array}{cc}
e_{L} & \overline{u_{R}^{r}}\end{array}\right]\]
Some right-chiral acting gauge fields, $\f{X^{1}}T_{1}^{R}+\f{X^{2}}T_{2}^{R}$,
are suggested for completeness, but left out as they're not part of
the standard model. The matrix of neutral and charged Higgs coefficients
is\[
\left[\begin{array}{cc}
\phi_{2}^{0} & \phi_{1}^{+}\\
\phi_{2}^{+} & \phi_{1}^{0}\end{array}\right]=\left[\begin{array}{cc}
\left(-i\phi^{4}+\phi^{7}\right) & \left(\phi^{5}-i\phi^{6}\right)\\
\left(\phi^{5}+i\phi^{6}\right) & \left(-i\phi^{4}-\phi^{7}\right)\end{array}\right]\]
 which comes from the definition of the Higgs vector field as\[
\phi=-\phi^{\psi}\Gamma_{\psi}\]
 and the spacetime vierbein as\[
\f{e}=\f{e^{\mu}}\Gamma_{\mu}P^{l}\]
 This definition of the Higgs and vierbein is compatible with $\phi\f{e}=\phi^{\psi}\f{e^{\mu}}\Gamma_{\mu\psi}P^{l}$,
but doesn't use all 16 degrees of freedom corresponding to the $\Gamma_{\mu\psi}P^{l}$
generators since there are only 4 corresponding to $\f{e^{\mu}}$
and 4 to $\phi^{\psi}$, or 8 if $\phi^{\psi}$ is allowed to be complex.
Also, since $\phi$ and $\f{e}$ multiply, one or the
other should be normalized -- letting the vierbein be free, the Higgs
vector is restricted to satisfy\[
\phi\cdot\phi=\phi^{\psi}\phi^{\chi}\eta_{\psi\chi}=-M^{2}\]

The curvature of this extended connection (the curvature of the standard
model and gravity) is\begin{align*}
\f{\tilde{\f{F}}}= & \f{\partial}\tilde{\f{A}}+\frac{1}{2}\tilde{\f{A}}\tilde{\f{A}}\\
= & \left(\f{\partial}\left(\phi\f{e}\right)+\left(\f{\omega}+\f{W}+\f{B}\right)\times\left(\phi\f{e}\right)\right)\\
 & +\left(\f{\partial}\f{\omega}+\frac{1}{2}\f{\omega}\f{\omega}+\frac{1}{2}\left(\phi\f{e}\right)\times\left(\phi\f{e}\right)\right)\\
 & +\left(\f{\partial}\f{W}+\frac{1}{2}\f{W}\f{W}\right)+\left(\f{\partial}\f{B}\right)+\left(\f{\partial}\f{G}+\frac{1}{2}\f{G}\f{G}\right)\\
 & +\left(\f{\partial}\psi+\frac{1}{2}\left(\phi\f{e}+\f{\omega}+\f{W}\right)\psi+\f{B}\times\psi-\frac{1}{2}\psi\f{G}\right)+\frac{1}{2}\psi\times\psi\end{align*}
 Most of these terms are familiar, but the Higgs terms require special
attention. The cosmological term is\[
\frac{1}{2}\left(\phi\f{e}\right)\times\left(\phi\f{e}\right)=\frac{1}{2}\phi\f{e}\phi\f{e}=-\frac{1}{2}\phi\phi\f{e}\f{e}=M^{2}\frac{1}{2}\f{e}\f{e}\]
 with cosmological constant equal to the normalization constant for
the Higgs, $\Lambda=M^{2}$. The first term, the Higgs extended torsion,
may be simplified since many of the objects commute, $\f{\omega}\times\phi=0$
and $\left(\f{W}+\f{B}\right)\times\f{e}=0$.
It breaks up into\[
\f{\f{T'}}=\left(\f{\partial}\phi+\left(\f{W}+\f{B}\right)\times\phi\right)\f{e}+\phi\left(\f{\partial}\f{e}+\f{\omega}\times\f{e}\right)\]
 which includes the gravitational torsion, $\f{\f{T}}=\f{\partial}\f{e}+\f{\omega}\times\f{e}$,
and the gauge covariant derivative of the Higgs multiplet. This is
rather unusual, as it relates the gravitational torsion to the weak
neutral gauge field, defined as $\f{Z}=\frac{1}{2}\left(\f{W^{3}}-\f{B}\right)$.
In terms of the representative matrices, and after multiplying by
the inverse vierbein, this looks like\[
\vg{e}\f{\f{T'}}\sim\left[\begin{array}{cc}
\f{\partial}+\frac{1}{4}\left(\vg{e}\f{\f{T}}\right)-\frac{i}{2}\f{W^{3}} & -\frac{i}{2}\f{W^{1}}-\frac{1}{2}\f{W^{2}}\\
-\frac{i}{2}\f{W^{1}}+\frac{1}{2}\f{W^{2}} & \f{\partial}+\frac{1}{4}\left(\vg{e}\f{\f{T}}\right)+\frac{i}{2}\f{W^{3}}\end{array}\right]\left[\begin{array}{cc}
\phi_{2}^{0} & \phi_{1}^{+}\\
\phi_{2}^{+} & \phi_{1}^{0}\end{array}\right]-\left[\begin{array}{cc}
\phi_{2}^{0} & \phi_{1}^{+}\\
\phi_{2}^{+} & \phi_{1}^{0}\end{array}\right]\left[\begin{array}{cc}
-\frac{i}{2}\f{B} & 0\\
0 & \frac{i}{2}\f{B}\end{array}\right]\]
 Since $\phi$ is normalized to $M$, it is reasonable to expand around
a constant value for the Higgs. If the Higgs coefficients, $\phi^{\psi}$,
are allowed to be complex, this value may be presumed to be\[
\left[\begin{array}{cc}
\phi_{2}^{0} & \phi_{1}^{+}\\
\phi_{2}^{+} & \phi_{1}^{0}\end{array}\right]=\left[\begin{array}{cc}
im_{2} & 0\\
0 & im_{1}\end{array}\right]\]
 with $m_{1}^{2}+m_{2}^{2}=M^{2}$. This then works as per the standard
Higgs mechanism to provide masses for the weak $\f{W}$
and $\f{Z}$ fields as well as Dirac masses for the
fermions, and spawns the massless photon field, $\f{A}=\frac{1}{2}\left(\f{W^{3}}+\f{B}\right)$.

The various charges can be read directly from the matrix representation
of the extended connection (\ref{eq:matrix}) by reading the coefficients
of $-i\f{W^{3}}$ and $-i\f{B}$ to the
left of any matrix element and the coefficient of $+i\f{B}$
below, since these act on the fermions from the left and right respectively
(\ref{eq:extF}). This produces the familiar table of charges:

\begin{center}\begin{tabular}{|c|c|c|c|c|c|c|c|c|c|c|c|c|}
\hline 
&
 $\nu_{L}^{e}$&
 $e_{L}$&
 $\nu_{R}^{e}$&
 $e_{R}$&
 $u_{L}$&
 $d_{L}$&
 $u_{R}$&
 $d_{R}$&
 $\phi^{0}$&
 $\phi^{+}$&
 $W^{1}$&
 $X^{1}$\tabularnewline
\hline
$W^{3}$&
 $1$&
 $-1$&
 $0$&
 $0$&
 $1$&
 $-1$&
 $0$&
 $0$&
 $-1$&
 $1$&
 $2$&
 $0$\tabularnewline
\hline
$B^{l}$&
 $0$&
 $0$&
 $1$&
 $-1$&
 $0$&
 $0$&
 $1$&
 $-1$&
 $0$&
 $0$&
 $0$&
 $1$\tabularnewline
\hline
$B^{r}$&
 $1$&
 $1$&
 $1$&
 $1$&
 $-\frac{1}{3}$&
 $-\frac{1}{3}$&
 $-\frac{1}{3}$&
 $-\frac{1}{3}$&
 $-1$&
 $-1$&
 $0$&
 $-1$\tabularnewline
\hline
$B=B^{l}-B^{r}$&
 $-1$&
 $-1$&
 $0$&
 $-2$&
 $\frac{1}{3}$&
 $\frac{1}{3}$&
 $\frac{4}{3}$&
 $-\frac{2}{3}$&
 $1$&
 $1$&
 $0$&
 $2$\tabularnewline
\hline
$A=\frac{1}{2}\left(W^{3}+B\right)$&
 $0$&
 $-1$&
 $0$&
 $-1$&
 $\frac{2}{3}$&
 $-\frac{1}{3}$&
 $\frac{2}{3}$&
 $-\frac{1}{3}$&
 $0$&
 $1$&
 $1$&
 $1$ \tabularnewline
\hline
\end{tabular}\end{center}

The Lagrangian for the standard model plus gravity may be written
in restricted BF form (\ref{eq:bfext}) with a nonlinear\[
\f{\f{\Phi}}=-\frac{1}{2}\f{\f{B^{\phi e+\omega}}}\gamma-\frac{1}{2}\f{\f{{*B^{W+B+G}}}}\]
 which includes the first term for gravity and a vierbein dependent
Hodge dual term for the Yang-Mills fields. As an alternative, it may
be interesting to consider a more unified Lagrangian with $\f{\f{\Phi}}=-\frac{1}{2}\f{\f{{*B}}}$,
which would imply the existence of an additional SKY Lagrangian term,
$\ub{L}\sim\frac{1}{2}\left\langle \f{\f{R}}\f{\f{{*R}}}\right\rangle $.

\subsection{Bivector $u(4)$ GUT}

Although that pretty much wraps it up, there are several significant
deficiencies worth mentioning. There is only one generation of fermions
represented, and since each complex quaternionic representative also
contains an anti-particle, these fermions are represented redundantly.
Also, the subgroup of $su(3)$ bivectors was picked out by hand with
the exclusion of other right handed generators. Furthermore, the massive
fermions all attain the same bare mass up to a single phase -- it
doesn't appear natural in this model to introduce separate Yukawa
couplings. Lastly, it seems somewhat ad-hoc to use the left/right
projector, $P^{l/r}$, in constructing the generators. These problems
may all be solved by stepping away from the familiar representations
of the standard model groups and considering what would happen if
all the generators were unprojected $\mathbb{C}l_{8}$ bivectors --
implying a potentially novel form of grand unification based on the
group $u(4)$. If the projectors are not used and the weak $su(2)$
generators are {}``stretched out,'' so they are {}``lopsided''
$8\times8$ instead of $4\times4$, and a set of bivector $su(3)$
generators are folded in, with a $u(1)$ gauge field added, the resulting
upper left and lower right quadrants of the representative matrix
could look something like:{\footnotesize \[
\left[\begin{array}{cccc}
\f{\omega^{L}}-3i\f{W^{3}}-3i\f{B} & -i\f{W^{1}}-\f{W^{2}} & -i\f{W^{1}}-\f{W^{2}}+\phi_{4}\f{e^{R}} & -i\f{W^{1}}-\f{W^{2}}+\phi_{2}\f{e^{R}}\\
-i\f{W^{1}}+\f{W^{2}} & \f{\omega^{L}}+i\f{W^{3}}-i\f{B}-i\f{G^{3}}-\frac{i}{\sqrt{3}}\f{G^{8}} & -i\f{G^{1}}-\f{G^{2}}+\phi_{3}\f{e^{R}} & -i\f{G^{4}}-\f{G^{5}}+\phi_{1}\f{e^{R}}\\
-i\f{W^{1}}+\f{W^{2}}-\phi_{1}\f{e^{L}} & -i\f{G^{1}}+\f{G^{2}}+\phi_{2}\f{e^{L}} & \f{\omega^{R}}+i\f{W^{3}}-i\f{B}+i\f{G^{3}}-\frac{i}{\sqrt{3}}\f{G^{8}} & -i\f{G^{6}}-\f{G^{7}}\\
-i\f{W^{1}}+\f{W^{2}}+\phi_{3}\f{e^{L}} & -i\f{G^{4}}+\f{G^{5}}-\phi_{4}\f{e^{L}} & -i\f{G^{6}}+\f{G^{7}} & \f{\omega^{R}}+i\f{W^{3}}-i\f{B}+\frac{2i}{\sqrt{3}}\f{G^{8}}\end{array}\right]\]
}This GUT model has likely been ruled out previously because these
$su(2)$ and $su(3)$ bivector generator subgroups of $su(4)$ do
not form proper subgroups. However, with this representation the offending
cross terms, $\f{W}\times\f{G}$, do not
fall in $\f{W}$ or $\f{G}$ but may stand
a chance of somehow being {}``absorbed'' by $\phi\f{e}$.
This model is very similar to, but slightly less extreme than, Tony
Smith's insightful model \cite{smith}. The generators (except possibly
for the $u(1)$) are all bivectors -- in fact exhausting the complete
set of 28. With this representation for the standard model gauge groups
and gravity, the upper right quadrant of three generations of fermions
might look something like:\[
\left[\begin{array}{cccc}
\left(\nu_{L}^{e}+\nu_{L}^{\mu}+\nu_{L}^{\tau}\right) & \left(u_{L}^{r}+u_{L}^{b}+u_{L}^{g}\right) & \left(s_{L}^{r}+s_{L}^{b}+s_{L}^{g}\right) & \left(b_{L}^{r}+b_{L}^{b}+b_{L}^{g}\right)\\
e_{L} & d_{L}^{r} & d_{L}^{b} & d_{L}^{g}\\
\mu_{R} & c_{R}^{r} & c_{R}^{b} & c_{R}^{g}\\
\tau_{R} & t_{R}^{r} & t_{R}^{b} & t_{R}^{g}\end{array}\right]\]
 However, the exact form would come from calculating the eigenvalues
(charges) and eigenvectors (fermions) corresponding to the standard
model gauge bivectors and labeling them accordingly. The mass eigenstates
would have to be independently calculated based on the Higgs vev's
and, through the power of wishful thinking, the CKM mixing matrices
established. One inevitable component, which will either make or break
this proposal, is a gravitational connection acting on spinor blocks
from the left and from the right. This idea is currently wild conjecture,
but provides a possible approach towards getting particle masses from
the structure of $\mathbb{C}l_{8}$-- although the true model describing
nature is likely to be a bit more complicated.

\section{Discussion}

This paper has progressed in small steps to construct a complete picture
of gravity and the standard model from the bottom up using basic elements
with as few mathematical abstractions as possible. It began and ended
with the description of a Clifford algebra as a graded Lie algebra,
which became the fiber over a four dimensional base manifold. The
connection and curvature of this bundle, along with an appropriately
restricted BF action, provided a complete description of General Relativity
in terms of Lie algebra valued differential forms, without use of
a metric. This {}``trick'' is equivalent to the MacDowell-Mansouri
method of getting GR from an $so(5)$ valued connection. Hamiltonian
dynamics were discussed, providing a possible connecting point with
the canonical approach to quantum gravity. Further tools and mathematical
elements were described just before they were needed. The matrix representation
of Clifford algebras was developed, as well as how spinor fields fit
in with these representations. The relevant BRST method produced spinor
fields with gauge operators acting on the left and right. These pieces
all came together, forming a complete picture of gravity and the standard
model as a single BRST extended connection. If this final picture
seems very simple, it has succeeded.

As a coherent picture, this work does have weaknesses. Everything
takes place purely on the level of {}``classical'' fields -- but
with an eye towards their use in a QFT via the methods of quantum
gravity, which must be applied in a truly complete model. The BRST
approach to deriving fermions from gauge symmetries, although a straightforward
application of standard techniques, may be hard to swallow. If this
method is unpalatable, it is perfectly acceptable to begin instead
with the picture of a fundamental fermionic field as a Clifford element
with gauge fields acting from the left and right in an appropriate
action. The model conjectured at the very end, based on the related
$u(4)$ GUT, is yet untested and should be treated with great skepticism
until further investigated. In a somewhat ironic twist, after arguing
in the beginning for the more natural description of the MM bivector
$so(5)$ model in terms of mixed grade $Cl_{1,3}$ vectors and bivectors,
this conjectured model is composed purely of bivector gauge fields.

Although the model stands on its own as a straightforward $\mathbb{C}l_{8}$
fiber bundle construction over four dimensional base, there are many
other compatible geometric descriptions. One alternative is to interpret
$\f{\tilde{A}}$ as the connection for a Cartan geometry
with Lie group $G$ -- with a Lie subgroup, $H$, formed by the generators
of elements other than $\f{e}$, and the spacetime {}``base''
formed by $G/H$. Another particularly appealing interpretation is
the Kaluza-Klein construction, with four compact dimensions implied
by the Higgs vector, $\phi=-\phi^{\psi}\Gamma_{\psi}$, and a corresponding
translation of the components of $\f{\tilde{A}}$ into
parts of a vielbein including this higher dimensional space. The model
may also be extended to encompass more traditional unification schemes,
such as using a ten dimensional Clifford algebra in a $so(10)$ GUT.
All of these geometric ideas should be developed further in the context
of the model described here, as they may provide valuable insights.

In conclusion, and in defense of its existence, this work has concentrated
on producing a clear and coherent unified picture rather than introducing
novel ideas in particular areas. The answer to the question of what
here is really {}``new'' is: {}``as little as possible.'' Rather,
several standard and non-standard pieces have been brought together
to form a unified whole describing the conventional standard model
and gravity as simply as possible.

\end{document}